\documentclass[twocolumn]{aastex61} 
\usepackage{csvsimple}

\newcommand{\acen}{\mbox{$\alpha~{\rm Cen}~$}} 
\newcommand{\alphacen}{\mbox{$\alpha \,{\rm Centauri}$}} 
\newcommand{\msun}{\mbox{$M_\odot$}} 
\newcommand{\lsun}{\mbox{$L_\odot$}}
\newcommand{\rsun}{\mbox{$R_\odot$}}
\newcommand{\mstar}{M$_{\star}~$}
\newcommand{\lstar}{\mbox{$L_\star~$}}
\newcommand{\rstar}{R$_{\star}~$}
\newcommand{\mjup}{M$_{\rm JUP}~$}
\newcommand{\mearth}{M$_\oplus~$}
\newcommand{\rearth}{R$_\oplus~$}
\newcommand{\msini}{$M \sin i~$}
\newcommand{\vsini}{$v \sin i~$}
\newcommand{\ms}{\mbox{m s$^{-1}~$}}
\newcommand{\cms}{\mbox{cm s$^{-1}~$}}
\newcommand{\kms}{\mbox{km s$^{-1}~$}}

\newcommand{\rhk}{$\log R'_{\rm HK}$}
\newcommand{\teff}{${\rm T_{eff}}~$}
\newcommand{\fe}{\rm [Fe/H]}
\newcommand{\logg}{${\rm \log g}~$}

\newcommand{\vmag}{$V_{\rm mag}$}

\newcommand{\mv}{M$_{\rm V}~$}
\newcommand{\kepler}{\mbox{\it Kepler }}

\received{July 1, 2017}
\submitjournal{ApJ}

\begin{document}

\title{Planet Detectability in the Alpha Centauri System}
\correspondingauthor{Lily Zhao}
\email{lily.zhao@yale.edu}

\author[0000-0002-3852-3590]{Lily Zhao}
\affil{Yale University, 52 Hillhouse, New Haven, CT 06511, USA}

\author[0000-0003-2221-0861]{Debra A. Fischer}
\affil{Yale University, 52 Hillhouse, New Haven, CT 06511, USA}

\author[0000-0002-9873-1471]{John Brewer}
\affil{Yale University, 52 Hillhouse, New Haven, CT 06511, USA}

\author[0000-0003-3261-0519]{Matt Giguere}
\affil{Yale University, 52 Hillhouse, New Haven, CT 06511, USA}

\author[0000-0002-0149-1302]{B{\'a}rbara Rojas-Ayala}
\affil{Departamento de Ciencias F\'{i}sicas, Universidad Andr\'{e}s Bello, Fern\'{a}ndez Concha 700 Edificio C1 Piso 3, 7591538 Santiago, Chile}

\begin{abstract}
We use more than a decade of radial velocity measurements for \acen A, B, and Proxima Centauri from HARPS, CHIRON, and UVES to identify the \msini and orbital periods of planets that could have been detected if they existed. At each point in a mass-period grid, we sample a simulated, Keplerian signal with the precision and cadence of existing data and assess the probability that the signal could have been produced by noise alone. Existing data places detection thresholds in the classically defined habitable zones at about \msini of 53 \mearth for \acen A, 8.4 \mearth for \acen B, and 0.47 \mearth for Proxima Centauri.  Additionally, we examine the impact of systematic errors, or ``red noise" in the data. A comparison of white- and red-noise simulations highlights quasi-periodic variability in the radial velocities that may be caused by systematic errors, photospheric velocity signals, or planetary signals. For example, the red-noise simulations show a peak above white-noise simulations at the period of Proxima Centauri b.  We also carry out a spectroscopic analysis of the chemical composition of the \alphacen\ stars.  The stars have super-solar metallicity with ratios of C/O and Mg/Si that are similar to the Sun, suggesting that any small planets in the \acen system may be compositionally similar to our terrestrial planets. Although the small projected separation of \acen A and B currently hampers extreme-precision radial velocity measurements, the angular separation is now increasing. By 2019, \acen A and B will be ideal targets for renewed Doppler planet surveys.
\end{abstract}


\section{Introduction} \label{sec:intro}
Over the past two decades, hundreds of exoplanets have been detected with the radial velocity technique, opening a new subfield of astronomy. In 2009, the NASA \kepler mission \citep{Borucki2011, Batalha2013} used the transit technique to dramatically advance our understanding of exoplanet architectures, especially for low-mass planets. \citet{Burke2015} used the Q1-Q16 \kepler catalog \citep{Mullally2015} with the \citet{Christiansen2015} pipeline completeness parameterization to assess planet occurrence rates for \kepler G and K dwarfs. For exoplanets with radii $0.75 \le R_{\rm planet} \le 2.5$ \rearth and orbital periods, $50 \le P_{\rm orb} \le 300$ days, they find an occurrence rate, $F_0 = 0.77$ planets per star, with an allowed range of $0.3 \le F_0 \le 1.9$. The \citet{Burke2015} \kepler data analysis suggests that most GK stars have rocky exoplanets and portends a bright future for the discovery of low-mass planets orbiting nearby GK stars with the radial velocity technique, once precision is improved.

At a distance of 1.3 parsecs, the three stars in the \alphacen\ system are our closest neighbors. The stars of the central, \acen AB binary system orbit each other with a semi-major axis of 24 AU and an eccentricity of 0.524 \citep{Pourbaix2016}. Though planets are now known to be common, there has been theoretical speculation about whether planets would form in such a close binary system \citep{Thebault2006, Thebault2008, Thebault2009}. Simulations have shown that if planets do form in this system \citep{Quintana2006,Quintana2007,Guedes2008}, there are regions where they can reside in dynamically stable orbits \citep{WiegertHolman1997,Quarles2016} around either \acen A and \acen B. Furthermore, approximately 20\% of known planets orbit stars that are components of binary star systems. Particularly interesting is the case of HD~196885~AB, a stellar binary system with a semi-major axis of 24 AU and an eccentricity of 0.409, similar to the orbit of \acen AB, with a known planet orbiting the primary star \citep{Correia2008, Fischer2009, Chauvin2007}. The case of HD~196885~Ab provides empirical evidence that the formation of planets is not precluded around \acen A or B. 

The third star, Proxima Centauri, is a smaller M dwarf and orbits this pair with a semi-major axis between 8,700 and greater than 10,000 AU \citep{WertheimerLaughlin2006,Kervella2017b}. The \alphacen\ system has long been a key target for Doppler planet searches from southern hemisphere observatories\citep{Murdoch1993,Endl2001,Dumusque2012,Endl2015}. While no planets have yet been discovered around \acen A or B \citep[c.f.][]{Dumusque2012, Hatzes2013, Rajpaul2016}, an Earth-mmass planet has been detected orbiting Proxima Centauri using the Doppler technique \citep{Anglada2016}. This recent discovery has increased interest in the system and the proximity of these stars is an enormous advantage for missions that aim to obtain images of any exoplanets. As human exploration ventures beyond our solar system, these closest stars will surely be our first destination.

In this work, we publish radial-velocity observations of \acen A and B, obtained at the Cerro Tololo Interamerican Observatory (CTIO) with the Echelle Spectrograph (ES) from 2008 - 2010 and the CTIO High Resolution (CHIRON) spectrograph. These data, together with archived data from the High Accuracy Radial Velocity Planet Searcher (HARPS) and the Ultraviolet and Visual Echelle Spectrograph (UVES) of \acen B and Proxima Centauri are used to test planet detectability and place constraints on the mass and orbital periods of putative planets that may remain undetected around these three stars.

\section{The alpha Centauri System} \label{sec:acen}
Alpha Centauri is a hierarchical triple-star system. The primary and secondary components, \acen A and B, are main-sequence stars with spectral types G2V and K1V, respectively, that are gravitationally bound in an eccentric orbit with a semi-major axis of about 24 AU. The two stars currently have an angular separation of about 5 arcseconds, which is not resolvable with the naked eye. Their combined brightness of -0.27 magnitudes makes \acen AB one of the brightest objects in the southern hemisphere. The third star in this system, \acen C or Proxima Centauri, was discovered in 1915 \citep{Innes1915} and is a relatively faint $V=11.1$ magnitude M6V dwarf at a projected angular separation of $2.2 \deg$ from \acen AB.

The recent astrometric analyses of \acen A \citep{vanLeeuwen2007, Pourbaix2016, Kervella2017a} yield an orbital parallax between 743 and 754 mas, corresponding to a distance of 1.33 to 1.35 pc away. The three stars in the \alphacen\ system are our closest stellar neighbors.

\subsection{Doppler Analysis}
Observations of \alphacen\ A and B were obtained with the 1.5-m telescope at the Cerro Tololo Interamerican Observatory (CTIO) in Chile. From 2008 - 2010, the refurbished Echelle Spectrometer (ES) was used to collect data.  The ES was located in the Coud{\'e} room; however no other attempt was made to stabilize the thermal environment of the spectrograph and both diurnal and seasonal variations resulted in temperature changes of several degrees in the Coud{\'e} room. Light from the telescope was coupled to this instrument with an optical fiber and a slit was positioned at the focus to set the resolution to $\sim 48,000$. However, the slit width was manually set with a micrometer and was not very precise, therefore, we expect that slight variations in the resolution occurred over time. 

In 2011, we replaced the ES with the CHIRON spectrograph \citep{Tokovinin2013}. This instrument was also placed in the Coud{\'e} room and the optical fiber was changed to an octagonal fiber to reduce modal noise in our spectra.  CHIRON was not in a vacuum enclosure, however the combination of thermal insulation and a thermally controlled space inside the Coud{\'e} room stabilized the temperature drifts to +/- 2 K.  There are four observing modes with CHIRON; for our observations of $\alpha$ Cen A and B we adopted a fixed-width slit at the focus of the optical fiber that provided an instrumental resolution of R $\sim$ 90,000 at the expense of a $\sim 30$\% light loss.  A small fraction of light was picked off from the light path inside the spectrograph and sent to a photomultiplier tube to determine the photon-weighted midpoint and correct for the barycentric velocity during our observations.

The ES and CHIRON both use an iodine cell to provide the wavelength solution and to model Doppler shifts \citep{Butler1996}. The iodine cell is inserted into the light path for all of the program observations where radial velocities will be measured. The forward modeling process that we use also requires high-SNR, high-resolution template observations and a very high resolution Fourier transform spectrum (FTS) of the iodine cell, obtained at the Pacific Northwest National Labs (PNNL) Environmental Molecular Sciences Laboratory (EMSL). Template observations are made without the iodine cell and are bracketed by several observations of bright, rapidly rotating B-stars through the iodine cell. The B-star observations are used to model the wavelength solution and the spectral line spread function (SLSF) of the instrument. The template observation is deconvolved with the SLSF, providing a higher resolution, iodine-free spectrum for modeling Doppler shifts. With the template observation, $T_s$ and the FTS iodine spectrum, $I_2$, the model is constructed as: 
\begin{equation}
    \left(T_s \times I_2\right) \ast SLSF 
\end{equation}
and a Levenberg-Marquardt least squares fitting is used to model the program observations. The error budget for the CHIRON radial velocity (RV) measurements accounts for instrumental errors (including variations in temperature, pressure, and vibrations), modal noise in the octagonal fiber, algorithm errors in the analysis, and the inclusion of velocity effects (granulation, spots, faculae) from the surface of the stars. For the \alphacen\ AB stars, flux contamination from the companion star turned out to be the most significant error source.

\subsection{Spectroscopic Analysis} \label{sec:stellar}
The stellar properties and chemical abundances of \acen A and B were determined by using the spectral synthesis modeling code, Spectroscopy Made Easy (SME), described in \citet{Brewer2016}, to analyze several iodine-free spectra obtained with the CHIRON spectrograph in 2012.  The stellar parameters that we derive, as well as some comparison data that represent the range of values from the published literature with available uncertainties, are listed in Table \ref{tab:stellar}.

Because we have analyzed 28 \acen A and B spectra, the rms of those spectroscopic parameters is one way to assess uncertainties. However, for all spectroscopic parameters, we find that the rms is too small to provide a plausible estimate of uncertainties. Instead, we adopt the more conservative model parameter uncertainties that were established using the same SME modeling technique for more than 1600 stars observed with the Keck HIRES spectrograph \citep{Brewer2016}. Following \citet{Brewer2016}, small empirical corrections were applied to the elemental abundances of \acen AB to account for slight systematic trends that occur as a function of temperature with our analysis method.

Our spectroscopic analysis yields an effective temperature of $5766 \pm 25$ K for \acen A, and $5218 \pm 25$ K for \acen B. The effective temperature for \acen A is consistent with the effective temperature measurement derived from angular-diameter measurements by \citet{Boyajian2013} and consistent with the G2V spectral classification \citep{Perryman1997, vanLeeuwen2007}.  The calculated effective temperature for \acen B is similarly consistent with the results of \citet{Boyajian2013} and the K1V spectral classification.

Both stars have a super-solar metallicity, \fe $= 0.22 \pm 0.03$ and $0.24 \pm 0.03$ for \acen A and B, respectively, consistent with other published values, e.g., \citet{Anderson2011}. We measure a ${\rm C/O}$ ratio of $0.47 \pm 0.05$ and ${\rm Mg/Si}$ of $1.05 \pm 0.03$ for \acen A, similar to the solar value. The results for \acen B are nearly identical with a ${\rm C/O}$ ratio of $0.49 \pm 0.05$ and a ${\rm Mg/Si}$ ratio of $1.05 \pm 0.03$, the same as \acen A. Because the ratios of abundances in stellar photospheres evolve slowly over main-sequence lifetimes \citep{Pinsonneault2001, Turcotte2002}, we can use the ${\rm C/O}$ and ${\rm Mg/Si}$ ratios as a proxy for disk compositions. \citet{BrewerFischer2016} showed that most stars have low C/O ratios, leaving the Mg/Si ratio important for regulating the geology of planetesimals. The implication is that any rocky planets forming around \acen A or B could have a composition and internal structure that may be similar to the solar system terrestrial planets.

The temperature and \fe\ for Proxima Centauri were derived from infrared K-band features in XSHOOTER spectra available from the ESO Public Archive. The observations were carried out in Period 92 using a slit width of 0.4$\arcsec$ ($R\sim$9100) and were reduced following the standard recipe described in the XSHOOTER pipeline manual\footnote{http://www.eso.org/sci/software/pipelines/} \citep{Vernet2011}. The wavelength calibration for the spectra was based on telluric lines, using a modified version for XSHOOTER data of the IDL-based code {\it xtellcor$\_$general} by \citet{Vacca2003}. The Proxima Cen spectra were convolved with a Gaussian kernel to degrade the resolution to $R\sim$2700, in order to use the Na I, Ca I, and H$_2$O-K2 indices calibrated to provide metallicity estimates for M~dwarf stars by \citet{Rojas-Ayala2012}. With this technique, we derive \teff $= 2879 \pm 50$ K and \fe\ $= 0.08 \pm 0.12$ and a spectral type of M5.5V. The metallicity for Proxima Cen is slightly lower than that of \acen A and B; however, the uncertainty in the Proxima Cen measurement is four times the uncertainty for \acen A or B. Proxima Centauri should share the same chemical composition as \acen A and B unless those stars had a significantly different accretion history than Proxima. 

\subsection{Isochrone Analysis}
Using spectroscopic parameters (\teff, [Fe/H], [Si/H]) and distance, we derive the best fit models to the Yale-Yonsei (${\rm Y^2}$) isochrones \citep{Demarque2004} to estimate the stellar mass, radius, and age for \acen A and B. Our stellar masses (listed in Table \ref{tab:stellar}) agree well with other published values \citep{Lundkvist2014, Pourbaix2016} and the radius is consistent with the angular-diameter measurement by \citet{Kervella2017a}. The isochrone-derived age for \acen A is $5.17^{+1.03}_{-0.97}$ Gyr, slightly older than the Sun and consistent with previous age estimates.  Our isochrone model for \acen B gives a younger age with large uncertainties, $2.53^{+3.12}_{-1.89}$ Gyr.  The posterior in the isochrone fit shows a peak at younger ages for \acen B that is ill-constrained by \logg and distance.  However, the ages for the two stars do agree within their uncertainties.

The log g, stellar mass, and radius of Proxima Cen were determined by adopting the age of $\sim$5 Gyr that we estimate for \acen AB, and interpolating the temperature onto a solar-metallicity isochrone for main-sequence, low-mass stars from \citet{Baraffe2015}. Because M dwarfs evolve very slowly after the pre-main-sequence phase, any errors in the adopted age of the star will not significantly affect the derived stellar model. The \citet{Baraffe2015} isochrones were only calculated for solar metallicity; therefore, the isochrone model parameters will not account for the slightly super-solar metallicity of Proxima Centauri. The isochrone model parameters for Proxima Cen are also compiled in Table \ref{tab:stellar}. 

\subsection{Chromospheric Activity}
The chromospheric activity of \acen A and B was monitored \citet{Henry1996} by measuring emission in the cores of the Ca II H \& K lines relative to continuum bandpasses (i.e., the ${S_{\rm HK}}$ values), scaled to the long-term Mount Wilson H \& K study \citep{Wilson1978, Vaughan1978, Duncan1991, Baliunas1995}. The ${S_{\rm HK}}$ values together with the $B-V$ of the star can then be transformed to \rhk, which is the fraction of bolometric luminosity from the lower chromosphere after subtracting off photospheric contributions \citep{Noyes1984b}. Using \rhk\ instead of ${S_{\rm HK}}$ allows for a straight-forward comparison of chromospheric activity that is independent of spectral type. 

Chromospheric activity provides a good way to estimate rotation periods and ages \citep{Noyes1984b}, even for older and more slowly rotating stars. Both \acen A and B are chromospherically quiet stars with estimated rotation periods of about 22 and 41 days, respectively \citep{Morel2000}. This is typical of stars that are about the age of the Sun. Coronal cycles have been measured at X-ray and UV wavelengths with periods of 19 and 8 years for \acen A and B, respectively \citep{Ayres2014, Ayres2015}. 

A recalibration of chromospheric activity-age relation and a calibration of stellar rotation to stellar age (gyrochronology) was carried out by \citet{Mamajek2008}. Their revised calibration returns activity and gyrochronology ages of $6.6$ and $4.4$ Gyr for \acen A and $5.2$ and $6.5$ Gyr for \acen B. \citet{Mamajek2008} estimate uncertainty in these ages of about $1.5$ Gyr for the activity calibration and $1.3$ Gyr for their gyrochronology technique.


\begin{deluxetable*}{lllll}[ht!]
\tabletypesize{\scriptsize}
\tablecaption{Stellar Parameters for the \acen Stars \label{tab:stellar}}
\tablenum{1}
\tablehead{
\colhead{Parameter} & \colhead{\acen A} & \colhead{\acen B} & \colhead{\acen C} & \colhead{Source} \\
}
\startdata
ID          & HIP71683, HD128620, GJ559A & HIP71681, HD128621, GJ559B & HIP70890, GJ551  &    \\
Spectral Type     & G2V      & K0V        & M5Ve       & \citet{Perryman1997}    \\
\vmag         &  $-0.01$    & $1.13$      & $11.1$     & \citet{Perryman1997}    \\
Parallax [mas]   & $754.81 \pm 4.11$ & $796.92 \pm 25.9$ & $771.64 \pm 2.6$ & \citet{vanLeeuwen2007}     \\
Parallax [mas]    & $743 \pm 1.3$ & $743 \pm 1.3$   & --        & \citet{Pourbaix2016}     \\
Parallax [mas]   & $747.17 \pm 0.61$ & $747 \pm 0.61$  & --        & \citet{Kervella2017a}    \\
$\mu$ RA [mas/yr]   & $-3679.25$   & $-3614.39$     & $-3775.75$    & \citet{Perryman1997}    \\
$\mu$ RA [mas/yr]   & --       & --         & $-3773.8 \pm 0.4$& \citet{Kervella2016}    \\
$\mu$ Dec [mas/yr]  & $473.67$    & $802.98$      & $765.65$    & \citet{Perryman1997}    \\
$\mu$ Dec [mas/yr]  &   --     &  --        & $770.5 \pm 2.0$ & \citet{Kervella2016}    \\
\mv          &  4.36     & 5.5        & 15.5   & using \citet{Kervella2017a} parallax \\
\mstar / \msun    & $1.10 \pm 0.03$ & $0.97 \pm 0.04$  & --        & \citet{Lundkvist2014}    \\
\mstar / \msun    & $1.13 \pm 0.007$ & $0.97 \pm 0.04$  & --        & \citet{Pourbaix2016}     \\
\mstar / \msun    & $1.1055 \pm 0.0039$ & $0.9373 \pm 0.0033$ & --      & \citet{Kervella2016}     \\
\mstar / \msun    & --       & --       & $0.1221 \pm 0.0022$ & \citet{Mann2015}      \\
\mstar / \msun    & --       & --        & $0.120 \pm 0.015$ & \citet{Anglada2016}      \\
\mstar / \msun    & $1.1 \pm 0.02$      & $0.91 \pm 0.02$       & $0.106 \pm 0.005$ & (this work)          \\
$\theta_{\rm LD}$ [mas] & $8.511 \pm 0.020$ & $6.001 \pm 0.0034$ & --       & \citet{Kervella2003}     \\
$\theta_{\rm LD}$ [mas] & --       & --       & $1.044 \pm 0.08$       & \citet{Segransan2003}     \\
\rstar / \rsun    & $1.22 \pm 0.01$ & $0.88 \pm 0.01$  & --        & \citet{Lundkvist2014}     \\
\rstar / \rsun    & $1.231 \pm 0.0036$ & $0.868 \pm 0.0052$ & --      & \citet{Pourbaix2016}     \\
\rstar / \rsun    & $1.2234 \pm 0.0053$ & $0.8632 \pm 0.00537$ & --     & \citet{Kervella2017a}     \\
\rstar / \rsun    & --       & --         & $0.1542 \pm 0.0045$ & \citet{Mann2015}      \\
\rstar / \rsun    & --       & --         & $0.141 \pm 0.02$ & \citet{Anglada2016}      \\
\rstar / \rsun    & $1.23 \pm 0.04$      & $0.84 \pm 0.02$       & $0.131 \pm 0.005$ & (this work)         \\
\lstar / \lsun    & $1.5159 \pm 0.0051$ & $0.5014 \pm 0.0017$ &       & \citet{Boyajian2013}    \\
\lstar / \lsun    & --       & --         & $0.00155 \pm 0.00005$ & \citet{Anglada2016}   \\
\lstar / \lsun    & --       & --         & $0.0011 \pm 0.0002$ & (this work)       \\
Age [Gyr]       & $5.17^{+1.03}_{-0.97}$   & $2.53^{+3.12}_{-1.89}$   & --        & (this work)          \\
Age [Gyr]       & $5.2 \pm 1$   & $4.5 \pm 1.2$   & --      & (isochrone) \citet{Boyajian2013} \\
Age [Gyr]       & $4.85 \pm 0.5$ & --         & --      & \citet{Thevenin2002}       \\
Age [Gyr]       & $6.6 \pm 1.6$  & $5.2 \pm 1.6$   & --      & (activity) \citet{Mamajek2008}  \\ 
Age [Gyr]       & $4.4 \pm 1.3$  & $6.5 \pm 1.3$   & --      & (gyro) \citet{Mamajek2008}   \\
\rhk         & --       & $-4.9 \pm 0.08$  & --        & \citet{Dumusque2012}     \\
\rhk         & $-5.002$    & $-4.923$      & --        & \citet{Henry1996}      \\
$P_{\rm rot}$ [days] & $22$      & $41$        & --        & \citet{Morel2000}    \\
\teff [K]     & $5793 \pm 7$  & $5232 \pm 9$    & --        & \citet{Boyajian2013}     \\
\teff [K]     & $5766 \pm 25$  & $5218 \pm 25$   & $2879 \pm 50$  & (this work)         \\ 
\vsini [\kms]     & $2.51 \pm 0.5$ & $1.9 \pm 0.5$   &         & (this work)    \\
\logg [\cms]    & $4.31 \pm 0.05$ & --         & --        & \citet{Lundkvist2014}    \\
\logg [\cms]    & $4.3117 \pm 0.0015$ & --       & --        & \citet{Kervella2017a}    \\
\logg [\cms]   & $4.27 \pm 0.05$ & $4.44 \pm 0.05$  & $5.23 \pm 0.01$ & (this work)         \\
\fe          & $0.20$     & $0.21$       & --        & \citet{Anderson2011}    \\
\fe          & $0.22 \pm 0.03$ & $0.24 \pm 0.03$  & $0.10 \pm 0.12$ & (this work)         \\
{\rm [C/H]}      & $0.19 \pm 0.03$ & $0.19 \pm 0.03$  & --        &  (this work)         \\
{\rm [O/H]}      & $0.25 \pm 0.04$ & $0.23 \pm 0.04$  & --        &  (this work)         \\
{\rm [Si/H]}     & $0.21 \pm 0.03$ & $0.20 \pm 0.03$  & --        &  (this work)        \\
{\rm [Mg/H]}     & $0.19 \pm 0.03$ & $0.19 \pm 0.03$  & --        &  (this work)         \\
{\rm [C/O]}      & $0.47 \pm 0.05$ & $0.49 \pm 0.05$  & --        &  (this work)         \\
{\rm [Mg/Si]}     & $1.05 \pm 0.03$ & $1.05 \pm 0.03$  & --        &  (this work)        \\
{\rm [Si/Fe]}     & $1.11 \pm 0.04$ & $1.04 \pm 0.03$  & --        &  (this work)         \\
\enddata
\end{deluxetable*}

\subsection{Stellar Ages} 
As described above, stellar ages can be estimated in several ways: from isochrone fitting, stellar activity, stellar rotation speed (gyrochronology), dynamical measurements of the visual binary orbit, or galactic kinematics. Furthermore, in the case of a binary star system, we expect that the stars are co-eval; both stars should yield an independent estimate for the age of the binary system. Taking the average of the stellar ages for both \acen A and B estimated from the ages tabulated in Table \ref{tab:stellar}, we calculate a weighted mean age for the \alphacen\ system of $5.03 \pm 0.34$ Gyr. 

\subsection{Stellar Orbits}
\begin{figure}[tp]
\centering
\includegraphics[width=0.45\textwidth]{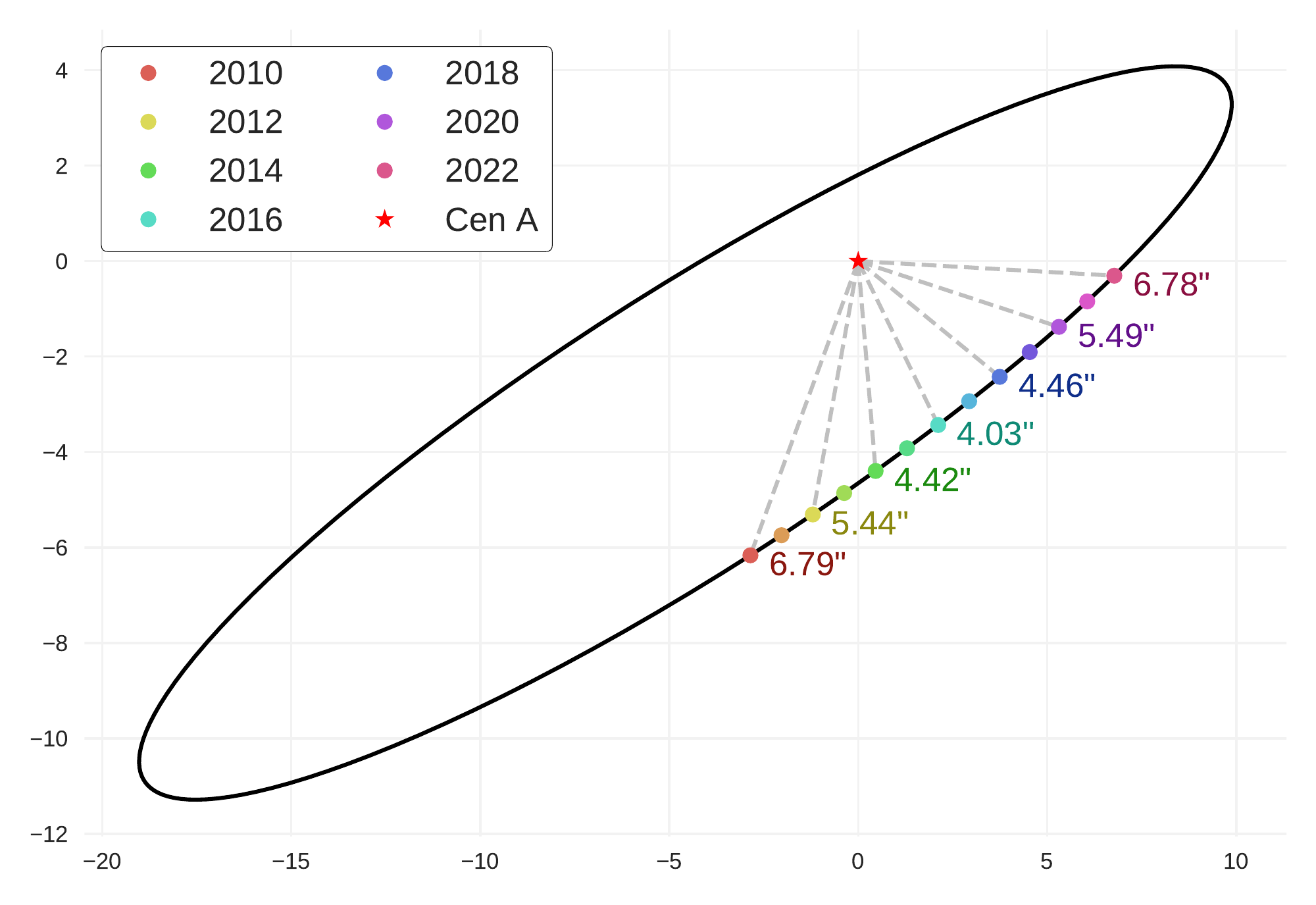}
\caption{Projected orbital plane of \acen A and B. The angular separation reaches a temporary minimum just under 4'' in 2017 and the angular separation begins to increase in 2018. By 2020, the separation exceeds 5\farcs5 and ground-based, radial-velocity searches can resume without suffering significant contamination from the companion star. }
\label{fig:f1}
\end{figure} 

\citet{Pourbaix1999} used published astrometry and radial velocity (RV) data from the European Southern Observatory Coud\'{e} Echelle Spectrograph to derive the orbital parameters for the \acen A and B stellar binary system. \citet{Pourbaix2016} refined this binary star orbit by supplementing their previous analysis with 11 years of high-precision RV measurements from the HARPS spectrograph and some additional astrometric data from the Washington Double Star Catalog \citep{Hartkopf2001}. They derive an orbital period of 79.91 $\pm$ 0.013 years and eccentricity of 0.524 $\pm$ 0.0011, and masses $M_{\rm A} = 1.133 \pm 0.005$ \msun\ and $M_{\rm B} = 0.972 \pm 0.0045$ \msun.

Proxima Centauri has a projected separation of $15,000 \pm 700$ AU from \acen AB and a relative velocity with respect to \acen AB of $0.53 \pm 0.14$ \kms \citep{WertheimerLaughlin2006}. \citet{WertheimerLaughlin2006} used Hipparcos kinematic information and carried out Monte Carlo simulations to determine the binding energy of Proxima Cen relative to \acen AB. They found a high probability that Proxima Cen is gravitationally bound and near apastron in a highly eccentric orbit. More recently, \citet{Kervella2017b} added published HARPS RV measurements and likewise concluded that Proxima Centauri is gravitationally bound to the \acen AB stars, traveling in an orbit with eccentricity of 0.50$^{+0.08}_{-0.09}$ with an orbital period of $\sim 550,000$ years.

\section{Exoplanet Searches} \label{sec:search}
The three stars in the \alphacen\ system have been targets of different precision, radial-velocity surveys to search for exoplanets from southern hemisphere observatories \citep{Endl2001, Dumusque2012, Endl2015}. In 2012, a planet was announced orbiting \acen B \citep{Dumusque2012} using data from the HARPS spectrograph. While that putative signal was later shown to be a sampling alias in the time series data \citep{Rajpaul2016}, \citet{Anglada2016} subsequently discovered a low-mass planet orbiting Proxima Centauri, a M5.5V star. The orbital period of Proxima Cen b is 11 days, which places this planet at the appropriate distance from its host star to fall within the habitable zone.  This detection was a record-breaking discovery because of the low mass of the planet, although the habitability of this world is now being debated. \citet{Airapetian2017} find that the planet orbiting Proxima Centauri will incur a significant atmospheric loss of oxygen and nitrogen in addition to a massive loss of hydrogen because of the high-energy flux from this relatively active M dwarf \citep{Airapetian2017}. 

There are several reasons why Doppler planet searches around the binary stars \acen A and B are well-motivated. The stars are bright, allowing for high cadence and signal-to-noise spectra. The declination of the stars is $-60$ degrees, close to a southern polar orbit, so that the observing season stretches between nine months and a year depending on the position of the observatory. Dynamical simulations \citep{WiegertHolman1997} show that any planets in the system are likely to be nearly aligned with the binary-star orbit; this implies that any RV amplitude would not be strongly attenuated by orbital inclination.

However, there are some challenges for planet detection, formation, and long-term stability around \acen A or B. One key concern is that the semi-major axis of the binary star orbit is only about 24 AU \citep{Pourbaix2016} and the orbital eccentricity of 0.524 means that the separation of the stars is only 16.3 AU at periastron passage. While \citet{WiegertHolman1997} demonstrate that any existing planets would be dynamically stable if they orbit within a few AU of either star, the close proximity of the stars has led to theoretical speculation about whether planets could have formed in the first place around \acen A or B \citep{Barbieri2002, Quintana2002, Quintana2006, Quintana2007, Thebault2008, Thebault2009}. Encouragingly, 20\% of detected exoplanets have been found in binary star systems orbiting one or the other star. An especially interesting case is the binary star HD~196885 AB. With a semi-major axis of 24 AU and an eccentricity of 0.409, this is a close analog of the \acen AB binary pair. HD~196885 A is known to host a gas-giant planet with \msini of $\sim 3$ \mjup\ and an orbital period of 3.69 years \citep{Correia2008, Fischer2009, Chauvin2007}.

Doppler surveys generally avoid binary stars with separations less than $\sim$5 arcseconds because additional RV errors can be incurred by flux contamination from the companion star. At the next periastron passage of \acen AB (May 2035) the projected separation of the two stars will be less than 2 arcseconds. However, with an orbital plane that is only 11 degrees from an edge-on configuration, the projected separation of \acen AB reached a secondary minimum of $\sim 4$ arcseconds in 2017. Figure \ref{fig:f1} shows the relative orbit of \acen B orbiting \acen A, projected onto the plane of the sky. Beginning in 2012, the angular separation between the two stars decreased to 5.44'' and flux contamination from the binary-star companion was observed in the radial-velocity measurements and was exacerbated on nights of poor seeing conditions.

For the CHIRON data, while there was code developed to scale the flux taking into account contamination, the improvement was insufficient for precision radial velocity measurements. The RVs listed in the \citet{Dumusque2012} paper were restricted to observations obtained through 2011 that had better than one arcsecond seeing. No HARPS radial velocities were published for 2012 because the seeing conditions were not adequate to avoid flux contamination during that year. \citet{Wittenmyer2014} and \citet{Bergmann2015} have presented methods for modeling flux-contaminated spectra to reach an rms of a few meters per second.  However, this more complex modeling does not reach sub-meter-per-second precision, the precision needed to contribute to the detection of planets with velocity semi-amplitudes less than one or two meters per second. The current small projected angular separation of \acen AB may force a hiatus in ground-based Doppler programs for this system until 2019 or 2020. 

\subsection{Constraints from Existing Data}
The existing Doppler planet searches allow us to place constraints on the mass-period parameter space where planets would have been detected if they existed. Conversely, we can see what type of planets would have escaped detection. 

Data from the Echelle Spectrograph (ES), CHIRON, HARPS, and UVES were compiled to constrain exoplanet detections for \acen A, B, and C (Proxima). Radial velocities of both \acen A and B were obtained by our team using the ES between 2008 - 2011 and the CHIRON spectrograph between 2011 - 2013 at the 1.5-m Cerro Tololo Interamerican Observatory (CTIO) in Chile. The HARPS spectrograph is located at the 3.6-m ESO La Silla telescope. HARPS radial velocities of \acen B were obtained between February 2008 and July 2011 and published by \citet{Dumusque2012}. We also use published RVs of Proxima Centauri from HARPS that span 2005 - 2016, and published RVs from 2010 - 2016 at the Ultraviolet and Visual Echelle Spectrograph (UVES) on the Very Large Telescope at Cerro Paranal in Chile \citep{Anglada2016}.  Both CHIRON and UVES are calibrated using the iodine cell technique while HARPS is calibrated using the simultaneous Thorium-Argon reference method \cite{Tokovinin2013,Anglada2016}.

The 1.5-m CTIO telescope is part of the Small to Moderate Aperture Research Telescopes (SMARTS) consortium. The ES was a recommissioned, fiber-fed spectrograph located at the 1.5-m CTIO telescope. The typical, single-shot precision of the ES was about 7 \ms. This spectrograph was replaced in 2011 with the CHIRON spectrograph, which was immediately upgraded and recommissioned in 2012 with new optical coatings, a new CCD, better temperature control, and octagonal fibers \citep{Tokovinin2013}. While the short term velocity rms reached 0.5 \ms\ for bright, single stars observed with CHIRON \citep{Tokovinin2013}, a serious short-coming for RV measurements of \acen A and B RV measurements is that the front end fiber feed was designed with a 2\farcs7 field of view to maximize the number of collected photons during poor seeing conditions on the 1.5-m telescope. When CHIRON was re-commissioned in 2012, the angular separation of \acen A and B was only 5\farcs5 and there was significant flux contamination from the companion star on nights when the seeing was worse than one arcsecond. By 2013, the angular separation of AB had decreased so the flux contamination increased and there were few nights when the rms of the RV measurements was less than three times the average scatter per night. We tested a new Doppler code that included a scaled flux from the companion star, as described by \citet{Bergmann2015}; however, we were only able to reach a single-shot precision of $\sim 15$ \ms\ from the flux contaminated spectra for \acen A and B.  We prefer to retain the original velocities, rather than velocities from our scaled flux analysis, because they more clearly identify nights with spectral contamination that should be rejected.

\begin{figure}[tp]
\centering
\includegraphics[width=0.45\textwidth]{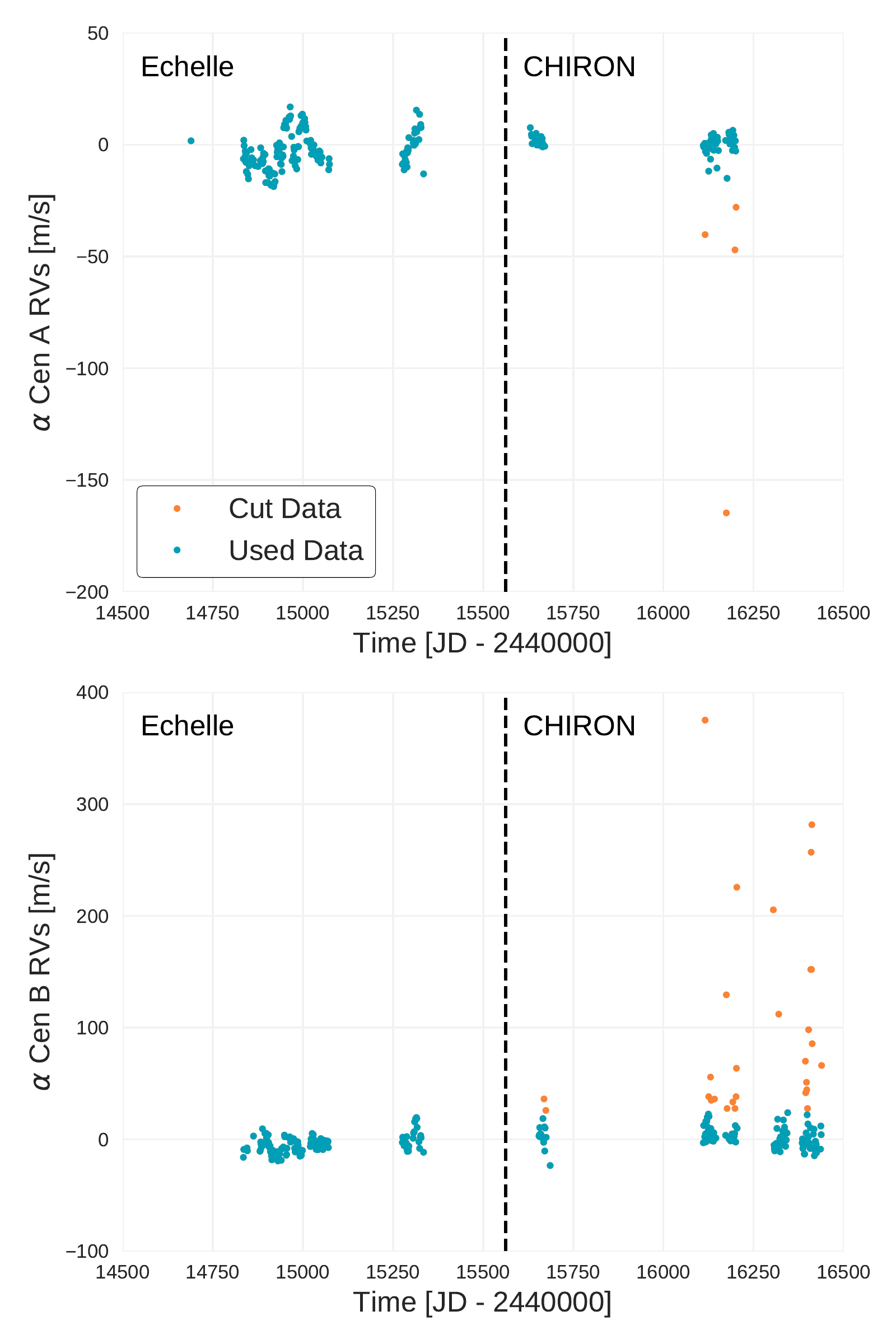}
\caption{Data of $\alpha$ Centauri A (top) and $\alpha$ Centauri B (bottom) taken at the CTIO Telescope from 2009 to 2012. The Echelle Spectrograph was switched to CHIRON in 2011, as shown by a dashed line on each graph.  Observations are binned by night.  Blue points represent the data points used in the simulations.  Observations falling more than 3$\sigma$ away from the average due to contamination were cut, shown here in orange.}
\label{fig:f-chiron}
\end{figure}

\begin{figure}[tp]
\centering
\includegraphics[width=0.45\textwidth]{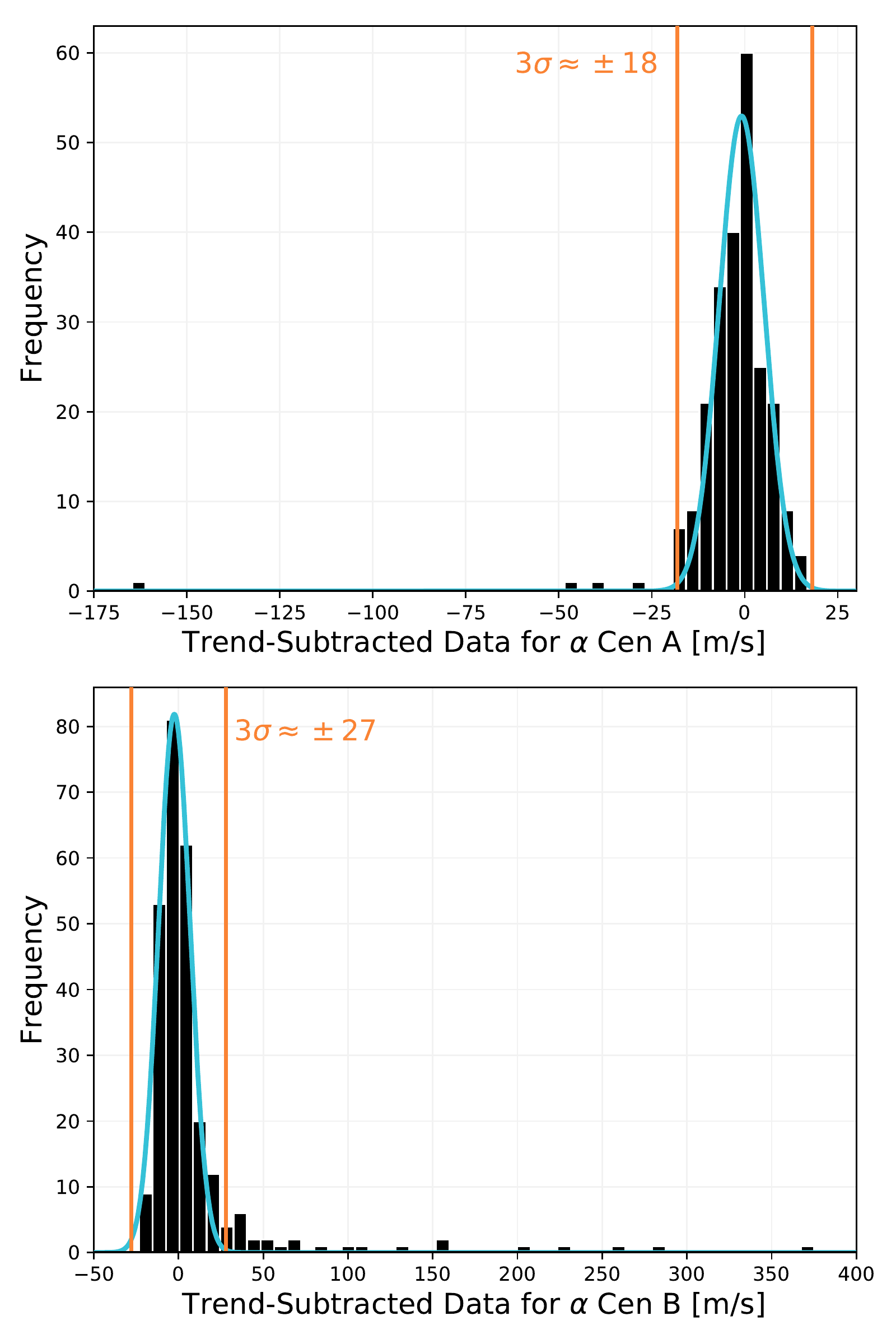}
\caption{A histogram of the de-trended radial velocity measurements for each night at the CTIO Telescope of \acen A (top) and \acen B (bottom).  Each histogram is fit to a Gaussian, shown in blue.  Nights where data fall more than 3$\sigma$ away (shown by red, vertical lines) most likely suffer contamination from the other star and are cut.  Retained nights are shown in blue in Figure \ref{fig:f-chiron}.}
\label{fig:f-cut}
\end{figure}

Figure \ref{fig:f-chiron} shows all of the binned RV measurements collected by the ES (left of the vertical dashed line) and CHIRON (right of the vertical dashed line) for \acen A (top panel) and \acen B (bottom panel). Flux contamination from the companion stars causes the velocities for \acen A to decrease (shifting toward the velocity of \acen B) and velocities for \acen B to increase. The effect of flux contamination is apparent in Figure \ref{fig:f-chiron}.  The nights with poor seeing conditions that resulted in flux contamination were excluded from the published HARPS data \citep{Dumusque2012}.  To eliminate nights at the CTIO with significant flux contamination, we determined an acceptable threshold for the measured contamination.  After subtracting out the binary trend from both data sets, the resulting RV measurements should be Gaussian distributed about zero.  However, contaminated data will lie far away from the mean.  We fit a Gaussian curve to the distribution of each data set and consider any data point more than 3$\sigma$ away from the distribution's mean as suffering from considerable contamination.  Figure \ref{fig:f-cut} shows the distribution (black), fitted Gaussian curve (blue), and subsequent cuts (orange).  The 3$\sigma$ cuts frame the bulk of the observations, thereby excluding only nights that deviate significantly from the mean.

Data that is retained and used in the simulations are shown in Figure \ref{fig:f-chiron} in blue while cut data is plotted in orange.  This contamination is visually obvious and increases with time following 2011 as the stars orbit closer and closer together (see Figure \ref{fig:f1}).  All velocities removed are skewed in the direction to be expected from contamination (e.g. down for \acen A and up for \acen B).  Additionally, this choice of cut vets more data points from the set of \acen B observations, which further suggests contamination since \acen A is the brighter star and would therefore cause more significant contamination in \acen B observations than the other way around.

In Table \ref{tab:rv}, we list the nightly-binned, radial-velocity measurements on nights where there was not significant flux contamination from the companion star for \alphacen\ A and B, resulting in 228 data points for \acen A and 241 points for \acen B. The uncertainty on our single measurements is of the order 5 \ms for ES data and 1.1 \ms for CHIRON data with regards to \acen A.  The uncertainty for \acen B observations is approximately 4.4 \ms for ES data and 1.2 \ms for CHIRON data.  Because the errors are not pure white noise, the error for each binned observations is taken to be the average of the formal errors of every point that night. The rms of the final, nightly-binned radial velocities is 7.2 \ms for \acen A over 4.13 years and 8.9 \ms for \acen B over 4.38 years.

\begin{deluxetable}{ccccc}[ht!]
\tabletypesize{\scriptsize}
\tablecaption{Relative, Binned RV Data from the Telescope \label{tab:rv}}
\tablenum{2}
\tablehead{
\colhead{Star} & \colhead{JD-2440000} & \colhead{RV \ms } & \colhead{Err \ms } & \colhead{Source}
}
\startdata
A   &   14689.5270  & -239.82  & 7.20  & Echelle \\
A   &   14834.8477  & -187.13  & 4.64  & Echelle \\ 
B   &   14834.8350  & 119.94  & 4.05  & Echelle \\
B   &   14835.8154  & 126.48  & 4.74  & Echelle \\
\enddata
\tablecomments{A stub of this table is provided in the printed version of this paper and the complete table is available in the online version of this paper}
\end{deluxetable}

\section{Simulations}
Using the cleaned and nightly-binned velocities for \acen A and B from the ES and CHIRON spectrographs at CTIO, the published HARPS velocities for \acen B, and the HARPS and UVES velocities for Proxima Cen, we carried out Monte Carlo simulations to assess whether planets of a given mass with orbital periods between 2 and 1000 days would have been detectable. The maximum orbital period of 1000 days was chosen because we expect that dynamical influences from the binary orbit of the \acen AB stars would destabilize orbits of putative planets beyond about 2 AU \citep{WiegertHolman1997}. We restricted the detectability simulations for Proxima Centauri (\acen C) to the same time baseline, searching for significant signals well beyond the habitable zone of the low-mass star. The minimum orbital period of 2 days is arbitrary, but avoids spurious 1-day sampling aliases in the CHIRON and HARPS data sets.

For the detectability simulations, we established a grid in planet mass and orbital period parameter space for each of the stars (\acen A, B, and Proxima Centauri) and injected a simulated Keplerian signal at each grid point, adopting stellar mass values from \citet{Pourbaix2016}. For simplicity, our simulations assume circular orbits and single-planet architectures. Grid points are spaced on a hybrid log-linear scale to adequately sample the parameter space.

The simulations reveal the detectable \msini of the planet. While our simulations test \msini rather than planet mass, planets are expected to inherit the $79^{\circ}$ orbital inclination of the binary star system \citep{WiegertHolman1997}.  Therefore, we expect that the \msini value is close to the true mass of any planets around either $\alpha$ Cen A or B.

The statistical significance of the injected signal was determined by  assuming the null hypothesis. In other words, we assess the probability that a signal of similar strength to our injected Keplerian signal would be produced by random errors in our data. Planets that are more massive or in closer orbits will produce stronger reflex velocities in the host stars that give rise to stronger, coherent signals.  These planets are more easily detected as their signals are harder to reproduce by noise alone. Our simulations test what strength of signal is necessary to overcome the inherent noise in the data and produce a coherent, detectable signal from a planet. We tested planet detectability in the presence of both white noise and the red noise present in the reported RVs.

\begin{figure}[t]
\centering
\includegraphics[width=.49\textwidth]{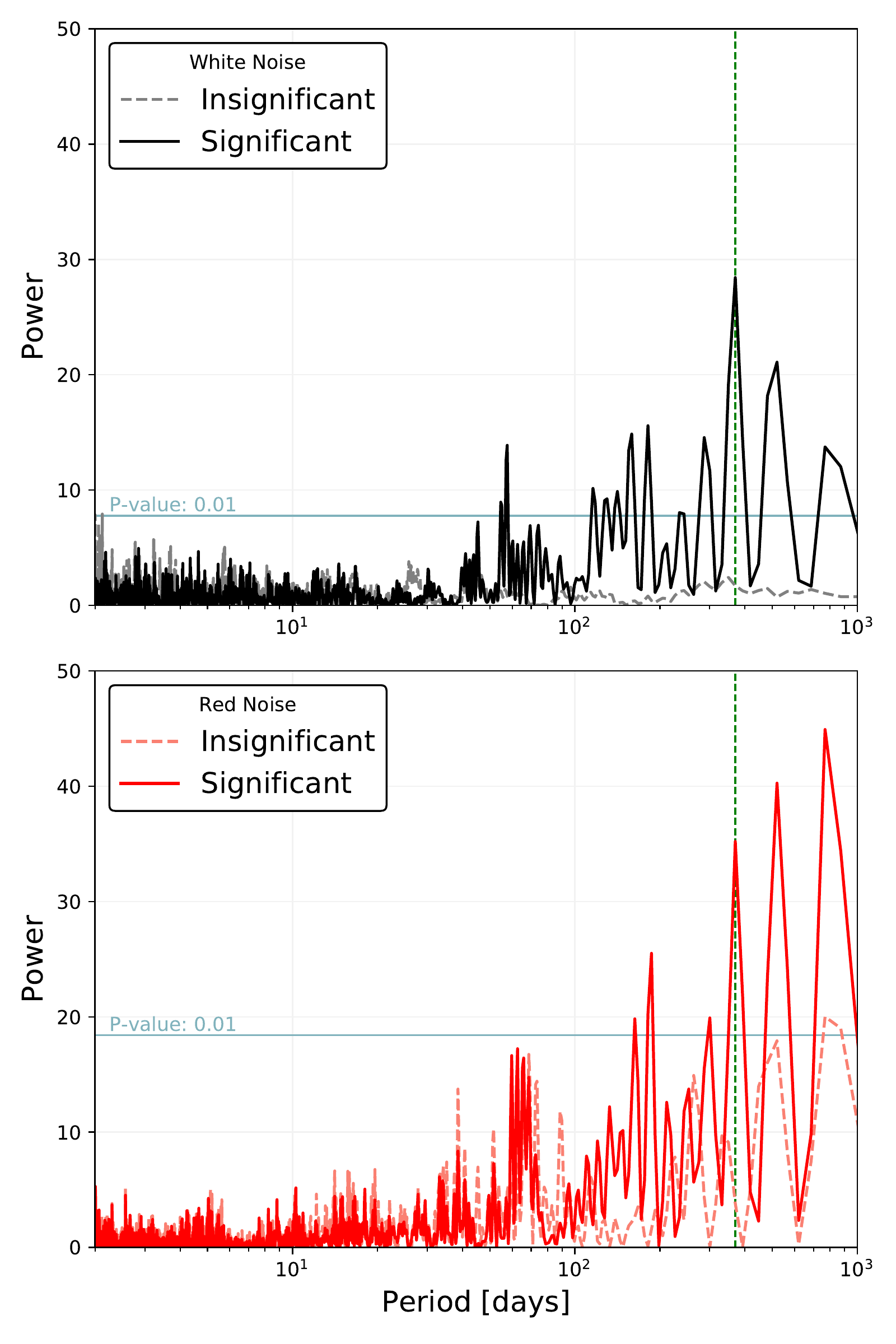}
\caption{Comparing generated Keplerian signals to noise.  An example periodogram of a significant detection (solid line) and an insignificant planetary signal (dashed-line) are given for white noise (black, top) and red noise (red, bottom).  On both graphs, a blue, horizontal line marks the peak height that is greater than the maximum peak height in 99\% of 1500 instances of pure noise.  This corresponds to a p-value of less than 0.01 for the signal.  Therefore, periodograms with peaks higher than this line are considered significant.  The vertical, green, dashed line on both graphs marks the period of the generated signal.}
\label{fig:f-pgram}
\end{figure}

\subsection{White-Noise Simulations}
\begin{figure*}[htp]
\centering
\includegraphics[width=1\textwidth]{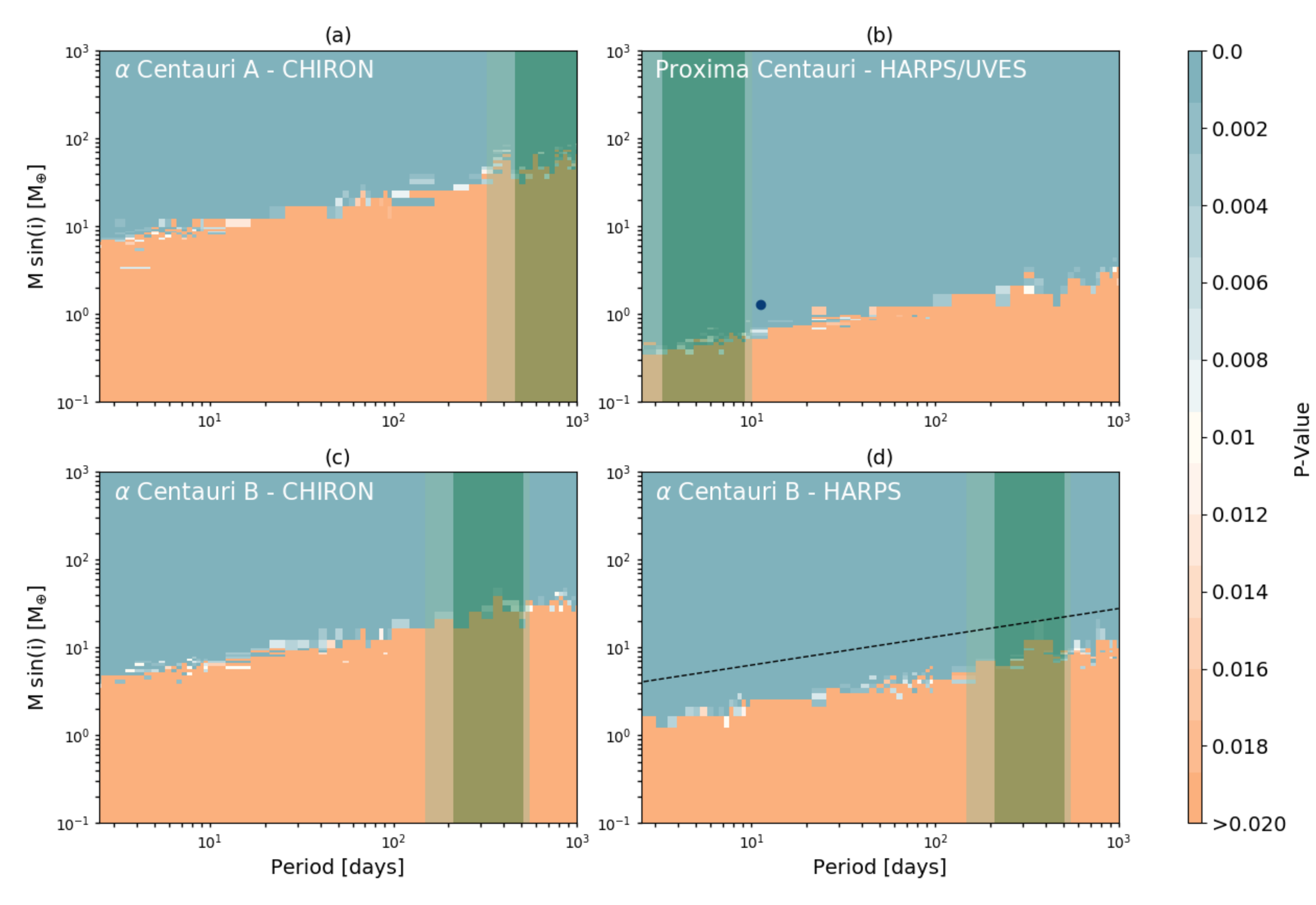}
\caption{White-noise simulations. Mass vs. period grids showing the significance at which a planet of such a mass and period would have been detected assuming only the reported errors for observations of (a) $\alpha$ Centauri A from ES and CHIRON, (b) Proxima Centauri from HARPS and UVES, (c) $\alpha$ Centauri B from ES and CHIRON, and (d) $\alpha$ Centauri B from HARPS. A p-value of less than 0.01 (indicated by shades of blue) is considered significant. Green vertical bands mark the conservative habitable zone where liquid water could persist for most of the stellar lifetime and the lighter green covers the optimistic habitable zone (as defined by \citet{Kopparapu2013}). A power law was fit to the detectability border of the \acen B ES and CHIRON data and is plotted on the \acen B HARPS grid as a dashed line.  The location of Proxima Cen b is indicated with a dot.}
\label{fig:f-wn}
\end{figure*}

\begin{figure*}[htp]
\centering
\includegraphics[width=1\textwidth]{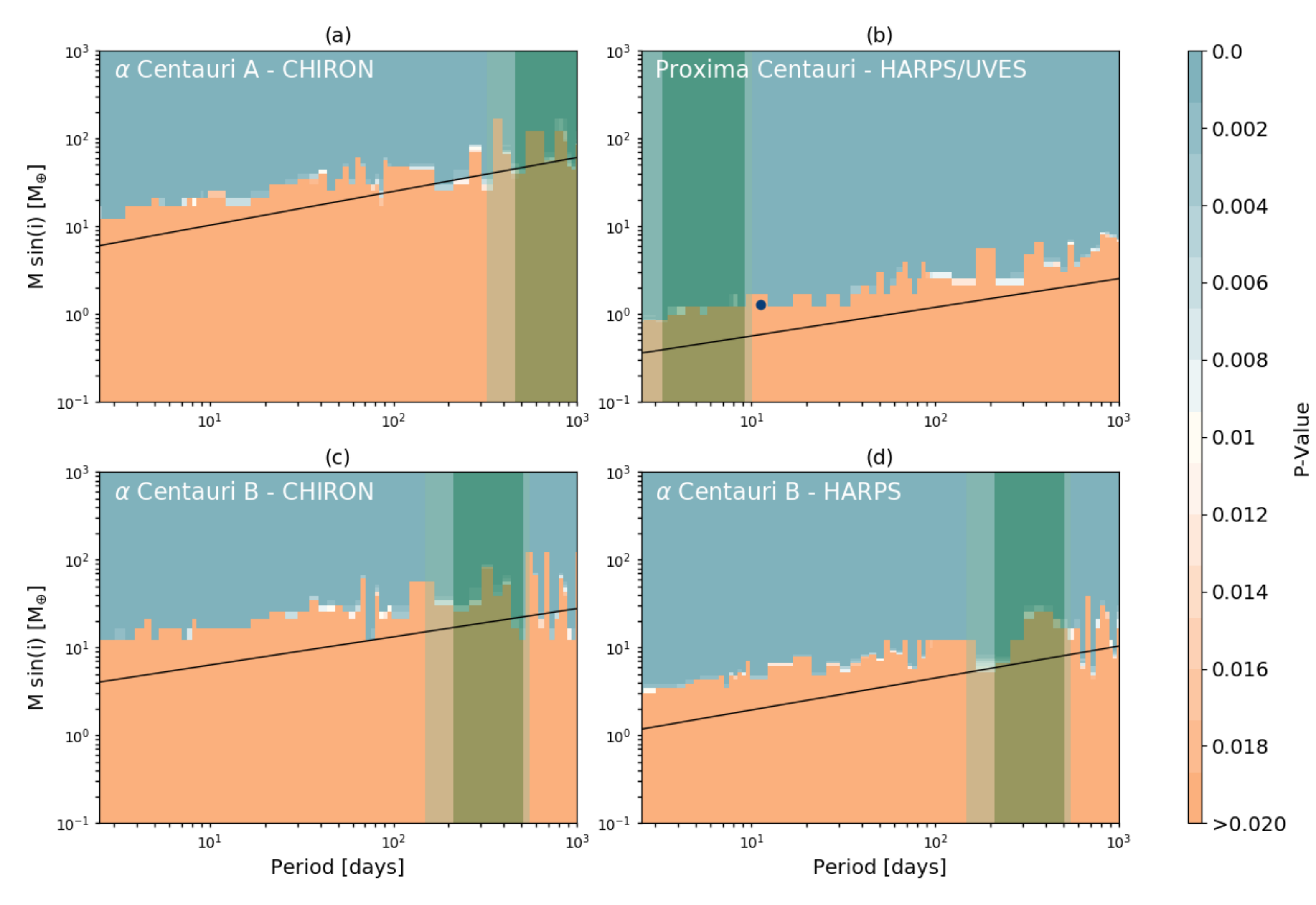}
\caption{Red-noise simulations. Mass vs. period grids showing the significance at which a planet of such a mass and period would have been detected assuming that the current data of (a) $\alpha$ Centauri A from ES and CHIRON, (b) Proxima Centauri from HARPS and UVES, (c) $\alpha$ Centauri B from ES and CHIRON, and (d) $\alpha$ Centauri B from HARPS is simply red noise. The color scale to p-value is the same as for Figure \ref{fig:f-wn}. A power law was fit to the detectability border given by the white-noise simulations and is plotted here as a black line. The orange parameter space indicates areas where planets could still remain undetected. The conservative and optimistic habitable zones are the same as Fig \ref{fig:f-wn}. Proxima Centauri b is indicated on subfigure (b) by a dot.}
\label{fig:f-rn}
\end{figure*}

We simulated Keplerian RVs with identical temporal sampling and error bars as the observed data sets, preserving any window functions in the observations. Our white-noise simulations assume that the radial velocity scatter is completely captured by white noise that is scaled to the quoted error bars. The simulated radial velocities for the white-noise simulations were created with a random draw from a Gaussian distribution that was scaled to the formal error at the time of each observation. The mean of the formal errors for the binned ES and CHIRON data is 0.48 \ms for $\alpha$ Centauri A and 0.51 \ms for $\alpha$ Centauri B (but, as we show later, the systematic errors in the ES and CHIRON data are significantly larger). The binned HARPS data of $\alpha$ Centauri B have a mean error of 1.0 \ms. The standard error for the combined, binned HARPS and UVES data of Proxima Centauri are on average 0.94 \ms. We generated 1500 sets of time-series, white-noise RV data. The simulated radial velocities were created by adding realizations of white noise to theoretical Keplerian models at each mass-period grid point.

Using a Lomb-Scargle periodogram \citep{Lomb1976,Scargle1982}, the periodogram power for the simulated Keplerian velocities was then compared to the periodogram power of the 1500 white noise data sets at each grid point. The data sets that are dominated by white noise produce a power spectrum with multiple low peaks at many periods, while a detectable Keplerian signal will produce a tall peak at the correct orbital period.  Examples of significant vs. insignificant periodograms are given in Figure \ref{fig:f-pgram} for both white noise (top) and red noise (bottom).

To decide whether the RVs with an injected Keplerian signal would be detectable with current observations, we calculate a p-value at each grid point. The p-value gives the probability that the injected radial-velocity data produces the same signal as only random noise. The p-value is defined as the fraction of comparisons where the white-noise simulations yield a greater maximum peak height than the simulated Keplerian signal. Planets producing significant signals will more consistently give stronger periodogram peaks, resulting in lower p-values. Larger p-values indicate that the signal produced by the planet has no more significance than white noise alone. We adopt an arbitrary but often used threshold p-value of 0.01, meaning that fewer than 1 of 100 white noise simulations produced a periodic signal that was stronger than a simulated Keplerian signal. \footnote{The purpose of this analysis is to assess the strength an injected Keplerian signal requires to produce a signal significantly distinct from what would be produced by pure noise in the current observations.  We consider each scenario individually and therefore do not make any adjustments to account for the multiple comparisons problem.}

Figure \ref{fig:f-wn} shows the white noise detectability simulations for \acen A, \acen B, and Proxima Cen. The p-values and color gradients are scaled so that a boundary appears where Keplerian signals yield a p-value of 0.01. Signals with lower p-values (above this boundary) would have likely been detected if they existed, while Keplerian signals with p-values greater than 0.01 would be buried in the white noise given the stated errors of the CHIRON and HARPS programs. The CHIRON and HARPS data sets for \acen B are kept separate as they are unique in their sampling and would lead to different aliasing as well as exhibit different instrumental errors.  Analyzing the two data sets separately allowed our results to capture these differences. Additionally, while combining the two data sets helps to push white-noise detection limits lower, the red-noise simulations suffer instead.  The CHIRON data, with more systematic errors, serve to reduce the sensitivity of the HARPS data rather than give better results. The upper right panel of Fig \ref{fig:f-wn} has a dot indicating the mass and period of Proxima Cen b \citep{Anglada2016}.
\pagebreak
\subsection{Red-Noise Simulations}
Our analysis using the white-noise simulations described above will not account for any systematic or quasi-periodic instrumental errors, analysis errors, photospheric jitter, or even actual planets. To investigate the impact of systematic errors or red noise sources, we treat the reported residual velocities from subtracting out the binary orbit from the observations as coherent noise. This is a worst-case scenario and we note that it is possible to improve detectability by de-correlating some of these noise sources using techniques like line bisector variations or FWHM variations to estimate photon noise \citep[e.g.][]{Dumusque2012, Rajpaul2016, Anglada2016}. 

These residual velocities are assumed to capture uncorrected observational errors, including instrumental errors and stellar jitter. The residual velocities would also contain any potential planetary signals. For this red noise simulation, we simply interpreted the residual velocities as pure red noise, and continued the same simulation described for white noise, adding Keplerian signals parameterized by each point of a mass-period grid. A comparison of the red and white noise analysis can be useful for highlighting possible planetary signals as well as quasi-periodic errors in our radial velocity data.

For the Monte Carlo, red noise simulations, the radial-velocity residuals were added to 1500 freshly generated white noise realizations and to the theoretical Keplerian signal at each point of a mass-period grid. The periodograms of the red-noise simulations now contain stronger power than the white-noise simulations, meaning the simulated Keplerian signal must generally have a larger amplitude to reach a p-value of 0.01 (see Figure \ref{fig:f-pgram}). Figure \ref{fig:f-rn} shows the red noise simulations for \acen A from CHIRON (upper left), Proxima Centauri (upper right), \acen B from CHIRON (lower left) and \acen B from HARPS (lower right). Solid, black lines on each plot show a power law that was fit to the detectability border from the white-noise simulations for each data set.

\section{Results}
\begin{figure}[htp]
\centering
\includegraphics[width=.49\textwidth]{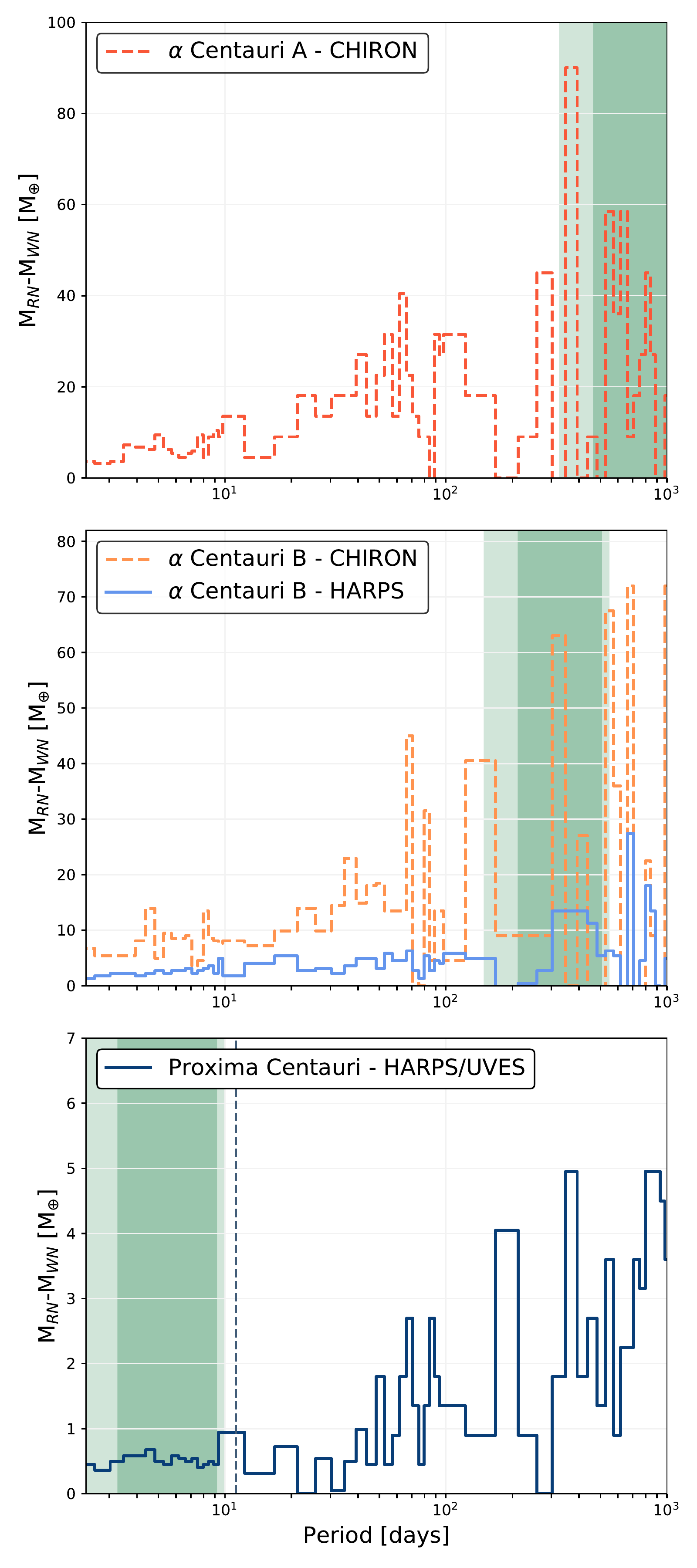}
\caption{Difference in detectability between the red-noise simulations and the white-noise simulations from subtracting the two detectabliity borders for $\alpha$ Centauri A (top), $\alpha$ Centauri B (middle) and $\alpha$ Centauri C (bottom). Peaks indicate periodicities in the residual radial velocities that could correspond to stellar noise, systematic errors, or even planetary signals. Negative differences are not shown. We assume negative values to arise from over-fitting the data or white noise alone and so hold no physical meaning. A blue, dashed line in the bottom, Proxima Centauri plot indicates the orbital period of Proxima Cen b.}
\label{fig:f-diff}
\end{figure}
Our white-noise simulations are summarized in Figure \ref{fig:f-wn}. Because both the original error bars and times of observation are retained, the white-noise simulations will preserve our ability to identify window functions in the sampling of our data. These simulations exclude planets in the conservative habitable zone of each star with a \msini of greater than 53$\pm$13 \mearth for \acen A, 8.4$\pm$1.5 \mearth for \acen B, and 0.47$\pm$0.08 \mearth for Proxima Centauri. However, this is an overly optimistic scenario. Doppler measurements are known to have contributions to the derived radial velocities that arise from instability of the instrument, errors in the analysis, and velocities in the stellar photosphere from spots, faculae, granulation, p-mode oscillations, or meridional flows \citep[e.g.,][]{Santos2000,Saar2000,Queloz2001,Wright2005,Lagrange2010,Meunier2010,Borgniet2015}. These velocities can obscure the Doppler signals that arise from orbiting exoplanets. The white noise simulations will not capture these noise sources because of an implicit assumption that the measurement errors are captured by the formal RV uncertainties. 

In contrast, our red-noise simulations, shown in Figure \ref{fig:f-rn}, represent a worst-case scenario. For these simulations, we assume that the time-series radial velocities contain only coherent noise. This noise is added directly to random white noise and the generated Keplerian signals, effectively preserving any temporal coherence in the noise.

In practice, radial velocities can be treated with Gaussian Process Regression \citep{Rajpaul2016} or decorrelated using line bisectors or the FWHM of the cross correlation function \citep{Dumusque2012} to mitigate the impact of non-Keplerian radial velocities on exoplanet detectability. Red noise has also been empirically modelled \citep{Tuomi2013} Therefore, our red-noise simulations slightly underestimate planet detectability. The white-noise and red-noise simulations together frame the mass-period boundary where existing Doppler surveys constrain the existence of planets orbiting the \alphacen\ stars.

The boundary between parameter space where planets would be detected or missed in the presence of white noise (Fig. \ref{fig:f-wn}) or red noise (Fig. \ref{fig:f-rn}) is approximately defined by the lowest mass at each period for which the p-value of the generated Keplerian RV exceeds our threshold of 0.01. This border for the white-noise simulations is both lower and smoother compared to the red-noise simulations. To more closely investigate these differences, we subtract the detectability border of the white-noise simulations, $M_{WN}$, from the detectability border of the red-noise simulations, $M_{RN}$. This difference is plotted as a function of period in Figure \ref{fig:f-diff} for \acen A (top), \acen B (middle), and for Proxima Cen (bottom). Peaks in this difference plot will occur due to quasi-periodic noise sources or planetary signals. Interesting to note is the peak at around 675-720 days that can be seen in both the CHIRON and HARPS observations of \acen B.  Additionally, similar peaks appear in both the \acen A and \acen B CHIRON data near 65, 150, and 575 days. The bottom plot in Figure \ref{fig:f-diff} includes a vertical, dashed line at the period of Proxima Centauri b, around which a clear peak can be seen.

\section{Discussion}
\subsection{Detectability}
We have carried out simulations to show how past Doppler surveys of the \alphacen\ stars constrain the probability of exoplanets over the mass-period parameter space shown in Figures \ref{fig:f-wn} and \ref{fig:f-rn}. While the Doppler technique can only derive $M \sin i$, rather than the true planet mass, the dynamical influences of the binary star system mean that any stable planets are likely to be nearly co-planar with the $79^\circ$ inclination of the stellar binary system \citep{WiegertHolman1997}. This suggests that the \msini is approximately the actual planet mass for prospective planets around \acen A or B. We show that Earth analogs could still exist around either \acen A or B and would not have been detected by the past decade of precision radial velocity searches. Continued, high-cadence, high-precision radial velocity observations could still reveal Earth-sized planets within this star system, even within the habitable zones of each of the three stars.

At each point in the parameter space of \msini and orbital period, we sample a Keplerian signal at the actual time of the observations with added white noise scaled to the errors to provide a baseline of planet detection space.  These simulations exclude planets within the conservative habitable zone of each planet with a \msini of greater than  53 \mearth for \acen A, 8.4 \mearth for \acen B, and 0.47 \mearth for Proxima Centauri on average.  This result for \acen B comes from the HARPS data set; the CHIRON data set excludes planets in the habitable zone of \acen B to greater than 23.5 \mearth.  We then repeat our analysis using the actual velocity scatter after subtracting the binary star orbit as ``red'' noise in addition to the white noise. We assess the probability that this signal could have been produced by noise alone by calculating a p-value, the fraction of comparisons where noise-only simulations yield greater periodogram power than the Keplerian signal at any grid point. The color scales of Figures \ref{fig:f-wn} and \ref{fig:f-rn} pivot around a p-value of 0.01, which would be marginally detectable. 

Both the white-noise and the red-noise simulations preserve the cadence of observations. Because observations are a discrete sampling of a continuous signal, aliases appear in periodograms that can be mistaken for true, astrophysical signals. These commonly correspond to periodicities of the sidereal year, sidereal day, solar day, and synodic month \citep{Dawson2010}. For example, reduced sensitivity can be seen in all four data sets presented in Figure \ref{fig:f-wn} around 300-400 days, which likely corresponds to the annual constraints the Earth's orbit around the Sun places on observations.

We subtract the white-noise mass-period boundary that occurs at a p-value of 0.01 from that same boundary in the red-noise simulations to highlight periodicities present in the residual radial velocities. Even after implementing the cuts described in section 3.1, we acknowledge that some of the remaining RV measurements may still be affected by small amounts of contamination, which effectively contributes to the red noise. The peaks apparent in Figure \ref{fig:f-diff} could correspond to quasi-periodic systematic errors, stellar jitter, or even planetary signals. For example, peaks that appear consistently in the CHIRON data for both \acen A and \acen B (e.g. at 65, 150, and 575 days), are indicative of instrumental or systematic errors since it is improbable that both stars will exhibit the same astrophysical velocity signals. Also potentially interesting are the periods where peaks in the CHIRON and HARPS data of \acen B align (e.g. at around 700 days). Because these peaks appear in observations from two different instruments with different data reduction pipelines, it seems unlikely that the same peaks would arise in both data sets from instrumental or systematic error; however, these peaks could still be the result of astrophysical velocity signals. In the case of Proxima Centauri, it is illustrative to note a distinct signal at the period of the recently discovered Proxima Cen b. A Keplerian signal would produce a red noise source in the velocities of that star; this peak is likely due to the signal produced by Proxima Cen b that is retained in the residual radial velocities.

Radial velocity precision approaching 10 centimeters per second will ultimately be needed to detect exoplanets with smaller masses and longer orbital periods in the yet to be probed parameter space around \acen A and B. There are several challenges for reaching such high RV precision. Some of the issues should be relatively straightforward to address. For example, the \emph{p}-mode oscillations of \acen A have a radial velocity amplitude of $1 - 3$ \ms \citep{Butler2004},which adds random scatter to radial velocities. The \emph{p}-mode amplitudes in \acen B are much weaker with a semi-amplitude of only 0.08 \ms \citep{Kjeldsen2005}; however, changing granulation patterns also introduce radial velocity scatter at the level of 0.6 \ms for both stars on timescales ranging from 15 minutes to several hours \citep{Dumusque2012, DelMoro2004}. Both \emph{p}-mode oscillations and granulation change on relatively short periods, allowing the observing strategy to be tailored to dramatically reduce these noise sources. For example, a series of exposures over ten minutes is sufficient to average over the high frequency \emph{p}-mode signals.

Additional challenges to higher RV precision include requirements of higher stability for next generation spectrographs (temperature and pressure stability), improved wavelength calibration, calibration of both CCD stitching and random pixel position errors, and mitigation of modal noise for multi-mode fibers \citep{Fischer2016}. It seems likely that ongoing efforts to address these engineering challenges will be successful. Techniques for modeling or decorrelating Doppler velocities that arise from stellar photospheres are less mature. Significant progress on disentangling stellar noise sources is required so that clean orbital velocities can be obtained. Currently, the two stars are separated by less than 5'', giving rise to cross contamination between the two stars and preventing high-precision, radial-velocity measurements. As the separation between \acen A and B begins to increase in 2019, radial velocity measurements will help to push constraints even lower and could ultimately lead to the discovery of Earth-like planets.

\acknowledgments
We gratefully acknowledge the anonymous referee whose insightful and specific comments served to significantly improve the paper. DAF gratefully acknowledges support from NASA NNH11ZDA001 and MJG thanks the NASA Earth and Space Fellowship 13-ASTRO13F-0011. LLZ gratefully acknowledges support from the NSF GRFP.  B.R-A acknowledges the support from CONICYT PAI/CONCURSO NACIONAL INSERCI\'ON EN LA ACADEMIA, CONVOCATORIA 2015 79150050. Based on observations collected at the European Organisation for Astronomical Research in the Southern Hemisphere under ESO programme 092.D-0300(A). Based on data obtained from the ESO Science Archive Facility under request number 275002.  We thank Tom Blake and the PNNL EMSL for obtaining the FTS scan of our iodine cell.  We also thank contributors to Matplotlib, the Python Programming Language, and the free and open-source community.

\facilities{Pacific Northwest National Labs (PNNL) Environmental Molecular Sciences Laboratory (EMSL), CTIO:1.5m}

\pagebreak
\bibliographystyle{aasjournal} 
\bibliography{main} 

\begin{thebibliography}{}
\expandafter\ifx\csname natexlab\endcsname\relax\def\natexlab#1{#1}\fi

\bibitem[{{Airapetian} {et~al.}(2017){Airapetian}, {Glocer}, {Khazanov},
  {Loyd}, {France}, {Sojka}, {Danchi}, \& {Liemohn}}]{Airapetian2017}
{Airapetian}, V.~S., {Glocer}, A., {Khazanov}, G.~V., {et~al.} 2017, \apjl,
  836, L3

\bibitem[{{Anderson} \& {Francis}(2011)}]{Anderson2011}
{Anderson}, E., \& {Francis}, C. 2011, VizieR Online Data Catalog, 5137

\bibitem[{{Anglada-Escud{\'e}} {et~al.}(2016){Anglada-Escud{\'e}}, {Amado},
  {Barnes}, {Berdi{\~n}as}, {Butler}, {Coleman}, {de La Cueva}, {Dreizler},
  {Endl}, {Giesers}, {Jeffers}, {Jenkins}, {Jones}, {Kiraga}, {K{\"u}rster},
  {L{\'o}pez-Gonz{\'a}lez}, {Marvin}, {Morales}, {Morin}, {Nelson}, {Ortiz},
  {Ofir}, {Paardekooper}, {Reiners}, {Rodr{\'{\i}}guez},
  {Rodr{\'{\i}}guez-L{\'o}pez}, {Sarmiento}, {Strachan}, {Tsapras}, {Tuomi}, \&
  {Zechmeister}}]{Anglada2016}
{Anglada-Escud{\'e}}, G., {Amado}, P.~J., {Barnes}, J., {et~al.} 2016, \nat,
  536, 437

\bibitem[{{Ayres}(2014)}]{Ayres2014}
{Ayres}, T.~R. 2014, \aj, 147, 59

\bibitem[{{Ayres}(2015)}]{Ayres2015}
---. 2015, \aj, 149, 58

\bibitem[{{Baliunas} {et~al.}(1995){Baliunas}, {Donahue}, {Soon}, {Horne},
  {Frazer}, {Woodard-Eklund}, {Bradford}, {Rao}, {Wilson}, {Zhang}, {Bennett},
  {Briggs}, {Carroll}, {Duncan}, {Figueroa}, {Lanning}, {Misch}, {Mueller},
  {Noyes}, {Poppe}, {Porter}, {Robinson}, {Russell}, {Shelton}, {Soyumer},
  {Vaughan}, \& {Whitney}}]{Baliunas1995}
{Baliunas}, S.~L., {Donahue}, R.~A., {Soon}, W.~H., {et~al.} 1995, \apj, 438,
  269

\bibitem[{{Baraffe} {et~al.}(2015){Baraffe}, {Homeier}, {Allard}, \&
  {Chabrier}}]{Baraffe2015}
{Baraffe}, I., {Homeier}, D., {Allard}, F., \& {Chabrier}, G. 2015, \aap, 577,
  A42

\bibitem[{{Barbieri} {et~al.}(2002){Barbieri}, {Marzari}, \&
  {Scholl}}]{Barbieri2002}
{Barbieri}, M., {Marzari}, F., \& {Scholl}, H. 2002, \aap, 396, 219

\bibitem[{{Batalha} {et~al.}(2013){Batalha}, {Rowe}, {Bryson}, {Barclay},
  {Burke}, {Caldwell}, {Christiansen}, {Mullally}, {Thompson}, {Brown},
  {Dupree}, {Fabrycky}, {Ford}, {Fortney}, {Gilliland}, {Isaacson}, {Latham},
  {Marcy}, {Quinn}, {Ragozzine}, {Shporer}, {Borucki}, {Ciardi}, {Gautier},
  {Haas}, {Jenkins}, {Koch}, {Lissauer}, {Rapin}, {Basri}, {Boss}, {Buchhave},
  {Carter}, {Charbonneau}, {Christensen-Dalsgaard}, {Clarke}, {Cochran},
  {Demory}, {Desert}, {Devore}, {Doyle}, {Esquerdo}, {Everett}, {Fressin},
  {Geary}, {Girouard}, {Gould}, {Hall}, {Holman}, {Howard}, {Howell},
  {Ibrahim}, {Kinemuchi}, {Kjeldsen}, {Klaus}, {Li}, {Lucas}, {Meibom},
  {Morris}, {Pr{\v s}a}, {Quintana}, {Sanderfer}, {Sasselov}, {Seader},
  {Smith}, {Steffen}, {Still}, {Stumpe}, {Tarter}, {Tenenbaum}, {Torres},
  {Twicken}, {Uddin}, {Van Cleve}, {Walkowicz}, \& {Welsh}}]{Batalha2013}
{Batalha}, N.~M., {Rowe}, J.~F., {Bryson}, S.~T., {et~al.} 2013, \apjs, 204, 24

\bibitem[{{Bergmann} {et~al.}(2015){Bergmann}, {Endl}, {Hearnshaw},
  {Wittenmyer}, \& {Wright}}]{Bergmann2015}
{Bergmann}, C., {Endl}, M., {Hearnshaw}, J.~B., {Wittenmyer}, R.~A., \&
  {Wright}, D.~J. 2015, International Journal of Astrobiology, 14, 173

\bibitem[{{Borgniet} {et~al.}(2015){Borgniet}, {Meunier}, \&
  {Lagrange}}]{Borgniet2015}
{Borgniet}, S., {Meunier}, N., \& {Lagrange}, A.-M. 2015, \aap, 581, A133

\bibitem[{{Borucki} {et~al.}(2011){Borucki}, {Koch}, {Basri}, {Batalha},
  {Boss}, {Brown}, {Caldwell}, {Christensen-Dalsgaard}, {Cochran}, {DeVore},
  {Dunham}, {Dupree}, {Gautier}, {Geary}, {Gilliland}, {Gould}, {Howell},
  {Jenkins}, {Kjeldsen}, {Latham}, {Lissauer}, {Marcy}, {Monet}, {Sasselov},
  {Tarter}, {Charbonneau}, {Doyle}, {Ford}, {Fortney}, {Holman}, {Seager},
  {Steffen}, {Welsh}, {Allen}, {Bryson}, {Buchhave}, {Chandrasekaran},
  {Christiansen}, {Ciardi}, {Clarke}, {Dotson}, {Endl}, {Fischer}, {Fressin},
  {Haas}, {Horch}, {Howard}, {Isaacson}, {Kolodziejczak}, {Li}, {MacQueen},
  {Meibom}, {Prsa}, {Quintana}, {Rowe}, {Sherry}, {Tenenbaum}, {Torres},
  {Twicken}, {Van Cleve}, {Walkowicz}, \& {Wu}}]{Borucki2011}
{Borucki}, W.~J., {Koch}, D.~G., {Basri}, G., {et~al.} 2011, \apj, 728, 117

\bibitem[{{Boyajian} {et~al.}(2013){Boyajian}, {von Braun}, {van Belle},
  {Farrington}, {Schaefer}, {Jones}, {White}, {McAlister}, {ten Brummelaar},
  {Ridgway}, {Gies}, {Sturmann}, {Sturmann}, {Turner}, {Goldfinger}, \&
  {Vargas}}]{Boyajian2013}
{Boyajian}, T.~S., {von Braun}, K., {van Belle}, G., {et~al.} 2013, \apj, 771,
  40

\bibitem[{{Brewer} \& {Fischer}(2016)}]{BrewerFischer2016}
{Brewer}, J.~M., \& {Fischer}, D.~A. 2016, \apj, 831, 20

\bibitem[{{Brewer} {et~al.}(2016){Brewer}, {Fischer}, {Valenti}, \&
  {Piskunov}}]{Brewer2016}
{Brewer}, J.~M., {Fischer}, D.~A., {Valenti}, J.~A., \& {Piskunov}, N. 2016,
  \apjs, 225, 32

\bibitem[{{Burke} {et~al.}(2015){Burke}, {Christiansen}, {Mullally}, {Seader},
  {Huber}, {Rowe}, {Coughlin}, {Thompson}, {Catanzarite}, {Clarke}, {Morton},
  {Caldwell}, {Bryson}, {Haas}, {Batalha}, {Jenkins}, {Tenenbaum}, {Twicken},
  {Li}, {Quintana}, {Barclay}, {Henze}, {Borucki}, {Howell}, \&
  {Still}}]{Burke2015}
{Burke}, C.~J., {Christiansen}, J.~L., {Mullally}, F., {et~al.} 2015, \apj,
  809, 8

\bibitem[{{Butler} {et~al.}(2004){Butler}, {Bedding}, {Kjeldsen}, {McCarthy},
  {O'Toole}, {Tinney}, {Marcy}, \& {Wright}}]{Butler2004}
{Butler}, R.~P., {Bedding}, T.~R., {Kjeldsen}, H., {et~al.} 2004, \apjl, 600,
  L75

\bibitem[{{Butler} {et~al.}(1996){Butler}, {Marcy}, {Williams}, {McCarthy},
  {Dosanjh}, \& {Vogt}}]{Butler1996}
{Butler}, R.~P., {Marcy}, G.~W., {Williams}, E., {et~al.} 1996, \pasp, 108, 500

\bibitem[{{Chauvin} {et~al.}(2007){Chauvin}, {Lagrange}, {Udry}, \&
  {Mayor}}]{Chauvin2007}
{Chauvin}, G., {Lagrange}, A.-M., {Udry}, S., \& {Mayor}, M. 2007, \aap, 475,
  723

\bibitem[{{Christiansen} {et~al.}(2015){Christiansen}, {Clarke}, {Burke},
  {Seader}, {Jenkins}, {Twicken}, {Catanzarite}, {Smith}, {Batalha}, {Haas},
  {Thompson}, {Campbell}, {Sabale}, \& {Kamal Uddin}}]{Christiansen2015}
{Christiansen}, J.~L., {Clarke}, B.~D., {Burke}, C.~J., {et~al.} 2015, \apj,
  810, 95

\bibitem[{{Correia} {et~al.}(2008){Correia}, {Udry}, {Mayor}, {Eggenberger},
  {Naef}, {Beuzit}, {Perrier}, {Queloz}, {Sivan}, {Pepe}, {Santos}, \&
  {S{\'e}gransan}}]{Correia2008}
{Correia}, A.~C.~M., {Udry}, S., {Mayor}, M., {et~al.} 2008, \aap, 479, 271

\bibitem[{{Dawson} \& {Fabrycky}(2010)}]{Dawson2010}
{Dawson}, R.~I., \& {Fabrycky}, D.~C. 2010, \apj, 722, 937

\bibitem[{{Del Moro}(2004)}]{DelMoro2004}
{Del Moro}, D. 2004, \aap, 428, 1007

\bibitem[{{Demarque} {et~al.}(2004){Demarque}, {Woo}, {Kim}, \&
  {Yi}}]{Demarque2004}
{Demarque}, P., {Woo}, J.-H., {Kim}, Y.-C., \& {Yi}, S.~K. 2004, \apjs, 155,
  667

\bibitem[{{Dumusque} {et~al.}(2012){Dumusque}, {Pepe}, {Lovis},
  {S{\'e}gransan}, {Sahlmann}, {Benz}, {Bouchy}, {Mayor}, {Queloz}, {Santos},
  \& {Udry}}]{Dumusque2012}
{Dumusque}, X., {Pepe}, F., {Lovis}, C., {et~al.} 2012, \nat, 491, 207

\bibitem[{{Duncan} {et~al.}(1991){Duncan}, {Vaughan}, {Wilson}, {Preston},
  {Frazer}, {Lanning}, {Misch}, {Mueller}, {Soyumer}, {Woodard}, {Baliunas},
  {Noyes}, {Hartmann}, {Porter}, {Zwaan}, {Middelkoop}, {Rutten}, \&
  {Mihalas}}]{Duncan1991}
{Duncan}, D.~K., {Vaughan}, A.~H., {Wilson}, O.~C., {et~al.} 1991, \apjs, 76,
  383

\bibitem[{{Endl} {et~al.}(2001){Endl}, {K{\"u}rster}, {Els}, {Hatzes}, \&
  {Cochran}}]{Endl2001}
{Endl}, M., {K{\"u}rster}, M., {Els}, S., {Hatzes}, A.~P., \& {Cochran}, W.~D.
  2001, \aap, 374, 675

\bibitem[{{Endl} {et~al.}(2015){Endl}, {Bergmann}, {Hearnshaw}, {Barnes},
  {Wittenmyer}, {Ramm}, {Kilmartin}, {Gunn}, \& {Brogt}}]{Endl2015}
{Endl}, M., {Bergmann}, C., {Hearnshaw}, J., {et~al.} 2015, International
  Journal of Astrobiology, 14, 305

\bibitem[{{Fischer} {et~al.}(2009){Fischer}, {Driscoll}, {Isaacson}, {Giguere},
  {Marcy}, {Valenti}, {Wright}, {Henry}, {Johnson}, {Howard}, {Peek}, \&
  {McCarthy}}]{Fischer2009}
{Fischer}, D., {Driscoll}, P., {Isaacson}, H., {et~al.} 2009, \apj, 703, 1545

\bibitem[{{Fischer} {et~al.}(2016){Fischer}, {Anglada-Escude}, {Arriagada},
  {Baluev}, {Bean}, {Bouchy}, {Buchhave}, {Carroll}, {Chakraborty}, {Crepp},
  {Dawson}, {Diddams}, {Dumusque}, {Eastman}, {Endl}, {Figueira}, {Ford},
  {Foreman-Mackey}, {Fournier}, {F{\H u}r{\'e}sz}, {Gaudi}, {Gregory},
  {Grundahl}, {Hatzes}, {H{\'e}brard}, {Herrero}, {Hogg}, {Howard}, {Johnson},
  {Jorden}, {Jurgenson}, {Latham}, {Laughlin}, {Loredo}, {Lovis}, {Mahadevan},
  {McCracken}, {Pepe}, {Perez}, {Phillips}, {Plavchan}, {Prato}, {Quirrenbach},
  {Reiners}, {Robertson}, {Santos}, {Sawyer}, {Segransan}, {Sozzetti},
  {Steinmetz}, {Szentgyorgyi}, {Udry}, {Valenti}, {Wang}, {Wittenmyer}, \&
  {Wright}}]{Fischer2016}
{Fischer}, D.~A., {Anglada-Escude}, G., {Arriagada}, P., {et~al.} 2016, \pasp,
  128, 066001

\bibitem[{{Guedes} {et~al.}(2008){Guedes}, {Rivera}, {Davis}, {Laughlin},
  {Quintana}, \& {Fischer}}]{Guedes2008}
{Guedes}, J.~M., {Rivera}, E.~J., {Davis}, E., {et~al.} 2008, \apj, 679, 1582

\bibitem[{{Hartkopf} {et~al.}(2001){Hartkopf}, {McAlister}, \&
  {Mason}}]{Hartkopf2001}
{Hartkopf}, W.~I., {McAlister}, H.~A., \& {Mason}, B.~D. 2001, \aj, 122, 3480

\bibitem[{{Hatzes}(2013)}]{Hatzes2013}
{Hatzes}, A.~P. 2013, \apj, 770, 133

\bibitem[{{Henry} {et~al.}(1996){Henry}, {Soderblom}, {Donahue}, \&
  {Baliunas}}]{Henry1996}
{Henry}, T.~J., {Soderblom}, D.~R., {Donahue}, R.~A., \& {Baliunas}, S.~L.
  1996, \aj, 111, doi:10.1086/117796

\bibitem[{{Innes}(1915)}]{Innes1915}
{Innes}, R.~T.~A. 1915, Circular of the Union Observatory Johannesburg, 30, 235

\bibitem[{{Kervella} {et~al.}(2017{\natexlab{a}}){Kervella}, {Bigot},
  {Gallenne}, \& {Th{\'e}venin}}]{Kervella2017a}
{Kervella}, P., {Bigot}, L., {Gallenne}, A., \& {Th{\'e}venin}, F.
  2017{\natexlab{a}}, \aap, 597, A137

\bibitem[{{Kervella} {et~al.}(2016){Kervella}, {Mignard}, {M{\'e}rand}, \&
  {Th{\'e}venin}}]{Kervella2016}
{Kervella}, P., {Mignard}, F., {M{\'e}rand}, A., \& {Th{\'e}venin}, F. 2016,
  \aap, 594, A107

\bibitem[{{Kervella} {et~al.}(2017{\natexlab{b}}){Kervella}, {Th{\'e}venin}, \&
  {Lovis}}]{Kervella2017b}
{Kervella}, P., {Th{\'e}venin}, F., \& {Lovis}, C. 2017{\natexlab{b}}, \aap,
  598, L7

\bibitem[{{Kervella} {et~al.}(2003){Kervella}, {Th{\'e}venin}, {S{\'e}gransan},
  {Berthomieu}, {Lopez}, {Morel}, \& {Provost}}]{Kervella2003}
{Kervella}, P., {Th{\'e}venin}, F., {S{\'e}gransan}, D., {et~al.} 2003, \aap,
  404, 1087

\bibitem[{{Kjeldsen} {et~al.}(2005){Kjeldsen}, {Bedding}, {Butler},
  {Christensen-Dalsgaard}, {Kiss}, {McCarthy}, {Marcy}, {Tinney}, \&
  {Wright}}]{Kjeldsen2005}
{Kjeldsen}, H., {Bedding}, T.~R., {Butler}, R.~P., {et~al.} 2005, \apj, 635,
  1281

\bibitem[{{Kopparapu} {et~al.}(2013){Kopparapu}, {Ramirez}, {Kasting}, {Eymet},
  {Robinson}, {Mahadevan}, {Terrien}, {Domagal-Goldman}, {Meadows}, \&
  {Deshpande}}]{Kopparapu2013}
{Kopparapu}, R.~K., {Ramirez}, R., {Kasting}, J.~F., {et~al.} 2013, \apj, 765,
  131

\bibitem[{{Lagrange} {et~al.}(2010){Lagrange}, {Desort}, \&
  {Meunier}}]{Lagrange2010}
{Lagrange}, A.-M., {Desort}, M., \& {Meunier}, N. 2010, \aap, 512, A38

\bibitem[{{Lomb}(1976)}]{Lomb1976}
{Lomb}, N.~R. 1976, \apss, 39, 447

\bibitem[{{Lundkvist} {et~al.}(2014){Lundkvist}, {Kjeldsen}, \& {Silva
  Aguirre}}]{Lundkvist2014}
{Lundkvist}, M., {Kjeldsen}, H., \& {Silva Aguirre}, V. 2014, \aap, 566, A82

\bibitem[{{Mamajek} \& {Hillenbrand}(2008)}]{Mamajek2008}
{Mamajek}, E.~E., \& {Hillenbrand}, L.~A. 2008, \apj, 687, 1264

\bibitem[{{Mann} {et~al.}(2015){Mann}, {Feiden}, {Gaidos}, {Boyajian}, \& {von
  Braun}}]{Mann2015}
{Mann}, A.~W., {Feiden}, G.~A., {Gaidos}, E., {Boyajian}, T., \& {von Braun},
  K. 2015, \apj, 804, 64

\bibitem[{{Meunier} {et~al.}(2010){Meunier}, {Desort}, \&
  {Lagrange}}]{Meunier2010}
{Meunier}, N., {Desort}, M., \& {Lagrange}, A.-M. 2010, \aap, 512, A39

\bibitem[{{Morel} {et~al.}(2000){Morel}, {Provost}, {Lebreton}, {Th{\'e}venin},
  \& {Berthomieu}}]{Morel2000}
{Morel}, P., {Provost}, J., {Lebreton}, Y., {Th{\'e}venin}, F., \&
  {Berthomieu}, G. 2000, \aap, 363, 675

\bibitem[{{Mullally} {et~al.}(2015){Mullally}, {Coughlin}, {Thompson}, {Rowe},
  {Burke}, {Latham}, {Batalha}, {Bryson}, {Christiansen}, {Henze}, {Ofir},
  {Quarles}, {Shporer}, {Van Eylen}, {Van Laerhoven}, {Shah}, {Wolfgang},
  {Chaplin}, {Xie}, {Akeson}, {Argabright}, {Bachtell}, {Barclay}, {Borucki},
  {Caldwell}, {Campbell}, {Catanzarite}, {Cochran}, {Duren}, {Fleming},
  {Fraquelli}, {Girouard}, {Haas}, {He{\l}miniak}, {Howell}, {Huber}, {Larson},
  {Gautier}, {Jenkins}, {Li}, {Lissauer}, {McArthur}, {Miller}, {Morris},
  {Patil-Sabale}, {Plavchan}, {Putnam}, {Quintana}, {Ramirez}, {Silva Aguirre},
  {Seader}, {Smith}, {Steffen}, {Stewart}, {Stober}, {Still}, {Tenenbaum},
  {Troeltzsch}, {Twicken}, \& {Zamudio}}]{Mullally2015}
{Mullally}, F., {Coughlin}, J.~L., {Thompson}, S.~E., {et~al.} 2015, \apjs,
  217, 31

\bibitem[{{Murdoch} {et~al.}(1993){Murdoch}, {Hearnshaw}, \&
  {Clark}}]{Murdoch1993}
{Murdoch}, K.~A., {Hearnshaw}, J.~B., \& {Clark}, M. 1993, \apj, 413, 349

\bibitem[{{Noyes} {et~al.}(1984){Noyes}, {Weiss}, \& {Vaughan}}]{Noyes1984b}
{Noyes}, R.~W., {Weiss}, N.~O., \& {Vaughan}, A.~H. 1984, \apj, 287, 769

\bibitem[{{Perryman} {et~al.}(1997){Perryman}, {Lindegren}, {Kovalevsky},
  {Hoeg}, {Bastian}, {Bernacca}, {Cr{\'e}z{\'e}}, {Donati}, {Grenon},
  {Grewing}, {van Leeuwen}, {van der Marel}, {Mignard}, {Murray}, {Le Poole},
  {Schrijver}, {Turon}, {Arenou}, {Froeschl{\'e}}, \&
  {Petersen}}]{Perryman1997}
{Perryman}, M.~A.~C., {Lindegren}, L., {Kovalevsky}, J., {et~al.} 1997, \aap,
  323, L49

\bibitem[{Pinsonneault {et~al.}(2001)Pinsonneault, DePoy, \&
  Coffee}]{Pinsonneault2001}
Pinsonneault, M.~H., DePoy, D.~L., \& Coffee, M. 2001, ApJ, 556, L59

\bibitem[{{Pourbaix} \& {Boffin}(2016)}]{Pourbaix2016}
{Pourbaix}, D., \& {Boffin}, H.~M.~J. 2016, \aap, 586, A90

\bibitem[{{Pourbaix} {et~al.}(1999){Pourbaix}, {Neuforge-Verheecke}, \&
  {Noels}}]{Pourbaix1999}
{Pourbaix}, D., {Neuforge-Verheecke}, C., \& {Noels}, A. 1999, \aap, 344, 172

\bibitem[{{Quarles} \& {Lissauer}(2016)}]{Quarles2016}
{Quarles}, B., \& {Lissauer}, J.~J. 2016, \aj, 151, 111

\bibitem[{{Queloz} {et~al.}(2001){Queloz}, {Henry}, {Sivan}, {Baliunas},
  {Beuzit}, {Donahue}, {Mayor}, {Naef}, {Perrier}, \& {Udry}}]{Queloz2001}
{Queloz}, D., {Henry}, G.~W., {Sivan}, J.~P., {et~al.} 2001, \aap, 379, 279

\bibitem[{{Quintana} {et~al.}(2007){Quintana}, {Adams}, {Lissauer}, \&
  {Chambers}}]{Quintana2007}
{Quintana}, E.~V., {Adams}, F.~C., {Lissauer}, J.~J., \& {Chambers}, J.~E.
  2007, \apj, 660, 807

\bibitem[{{Quintana} \& {Lissauer}(2006)}]{Quintana2006}
{Quintana}, E.~V., \& {Lissauer}, J.~J. 2006, \icarus, 185, 1

\bibitem[{{Quintana} {et~al.}(2002){Quintana}, {Lissauer}, {Chambers}, \&
  {Duncan}}]{Quintana2002}
{Quintana}, E.~V., {Lissauer}, J.~J., {Chambers}, J.~E., \& {Duncan}, M.~J.
  2002, \apj, 576, 982

\bibitem[{{Rajpaul} {et~al.}(2016){Rajpaul}, {Aigrain}, \&
  {Roberts}}]{Rajpaul2016}
{Rajpaul}, V., {Aigrain}, S., \& {Roberts}, S. 2016, \mnras, 456, L6

\bibitem[{{Rojas-Ayala} {et~al.}(2012){Rojas-Ayala}, {Covey}, {Muirhead}, \&
  {Lloyd}}]{Rojas-Ayala2012}
{Rojas-Ayala}, B., {Covey}, K.~R., {Muirhead}, P.~S., \& {Lloyd}, J.~P. 2012,
  \apj, 748, 93

\bibitem[{{Saar} \& {Fischer}(2000)}]{Saar2000}
{Saar}, S.~H., \& {Fischer}, D. 2000, \apjl, 534, L105

\bibitem[{{Santos} {et~al.}(2000){Santos}, {Mayor}, {Naef}, {Pepe}, {Queloz},
  {Udry}, \& {Blecha}}]{Santos2000}
{Santos}, N.~C., {Mayor}, M., {Naef}, D., {et~al.} 2000, \aap, 361, 265

\bibitem[{{Scargle}(1982)}]{Scargle1982}
{Scargle}, J.~D. 1982, \apj, 263, 835

\bibitem[{{S{\'e}gransan} {et~al.}(2003){S{\'e}gransan}, {Kervella},
  {Forveille}, \& {Queloz}}]{Segransan2003}
{S{\'e}gransan}, D., {Kervella}, P., {Forveille}, T., \& {Queloz}, D. 2003,
  \aap, 397, L5

\bibitem[{{Th\'{e}bault} {et~al.}(2006){Th\'{e}bault}, {Marzari}, \&
  {Scholl}}]{Thebault2006}
{Th\'{e}bault}, P., {Marzari}, F., \& {Scholl}, H. 2006, Icarus, 183, 193

\bibitem[{{Th\'{e}bault} {et~al.}(2008){Th\'{e}bault}, {Marzari}, \&
  {Scholl}}]{Thebault2008}
---. 2008, \mnras, 388, 1528

\bibitem[{{Th\'{e}bault} {et~al.}(2009){Th\'{e}bault}, {Marzari}, \&
  {Scholl}}]{Thebault2009}
---. 2009, \mnras, 393, L21

\bibitem[{{Th{\'e}venin} {et~al.}(2002){Th{\'e}venin}, {Provost}, {Morel},
  {Berthomieu}, {Bouchy}, \& {Carrier}}]{Thevenin2002}
{Th{\'e}venin}, F., {Provost}, J., {Morel}, P., {et~al.} 2002, \aap, 392, L9

\bibitem[{{Tokovinin} {et~al.}(2013){Tokovinin}, {Fischer}, {Bonati},
  {Giguere}, {Moore}, {Schwab}, {Spronck}, \& {Szymkowiak}}]{Tokovinin2013}
{Tokovinin}, A., {Fischer}, D.~A., {Bonati}, M., {et~al.} 2013, \pasp, 125,
  1336

\bibitem[{{Tuomi} {et~al.}(2013){Tuomi}, {Jones}, {Jenkins}, {Tinney},
  {Butler}, {Vogt}, {Barnes}, {Wittenmyer}, {O'Toole}, {Horner}, {Bailey},
  {Carter}, {Wright}, {Salter}, \& {Pinfield}}]{Tuomi2013}
{Tuomi}, M., {Jones}, H.~R.~A., {Jenkins}, J.~S., {et~al.} 2013, \aap, 551, A79

\bibitem[{Turcotte \& Wimmer-Schweingruber(2002)}]{Turcotte2002}
Turcotte, S., \& Wimmer-Schweingruber, R.~F. 2002, J Geophys Res-Space, 107,
  1442

\bibitem[{{Vacca} {et~al.}(2003){Vacca}, {Cushing}, \& {Rayner}}]{Vacca2003}
{Vacca}, W.~D., {Cushing}, M.~C., \& {Rayner}, J.~T. 2003, \pasp, 115, 389

\bibitem[{{van Leeuwen}(2007)}]{vanLeeuwen2007}
{van Leeuwen}, F. 2007, \aap, 474, 653

\bibitem[{{Vaughan} {et~al.}(1978){Vaughan}, {Preston}, \&
  {Wilson}}]{Vaughan1978}
{Vaughan}, A.~H., {Preston}, G.~W., \& {Wilson}, O.~C. 1978, \pasp, 90, 267

\bibitem[{{Vernet} {et~al.}(2011){Vernet}, {Dekker}, {D'Odorico}, {Kaper},
  {Kjaergaard}, {Hammer}, {Randich}, {Zerbi}, {Groot}, {Hjorth}, {Guinouard},
  {Navarro}, {Adolfse}, {Albers}, {Amans}, {Andersen}, {Andersen}, {Binetruy},
  {Bristow}, {Castillo}, {Chemla}, {Christensen}, {Conconi}, {Conzelmann},
  {Dam}, {de Caprio}, {de Ugarte Postigo}, {Delabre}, {di Marcantonio},
  {Downing}, {Elswijk}, {Finger}, {Fischer}, {Flores}, {Fran{\c c}ois},
  {Goldoni}, {Guglielmi}, {Haigron}, {Hanenburg}, {Hendriks}, {Horrobin},
  {Horville}, {Jessen}, {Kerber}, {Kern}, {Kiekebusch}, {Kleszcz}, {Klougart},
  {Kragt}, {Larsen}, {Lizon}, {Lucuix}, {Mainieri}, {Manuputy}, {Martayan},
  {Mason}, {Mazzoleni}, {Michaelsen}, {Modigliani}, {Moehler}, {M{\o}ller},
  {Norup S{\o}rensen}, {N{\o}rregaard}, {P{\'e}roux}, {Patat}, {Pena}, {Pragt},
  {Reinero}, {Rigal}, {Riva}, {Roelfsema}, {Royer}, {Sacco}, {Santin},
  {Schoenmaker}, {Spano}, {Sweers}, {Ter Horst}, {Tintori}, {Tromp}, {van
  Dael}, {van der Vliet}, {Venema}, {Vidali}, {Vinther}, {Vola}, {Winters},
  {Wistisen}, {Wulterkens}, \& {Zacchei}}]{Vernet2011}
{Vernet}, J., {Dekker}, H., {D'Odorico}, S., {et~al.} 2011, \aap, 536, A105

\bibitem[{{Wertheimer} \& {Laughlin}(2006)}]{WertheimerLaughlin2006}
{Wertheimer}, J.~G., \& {Laughlin}, G. 2006, \aj, 132, 1995

\bibitem[{{Wiegert} \& {Holman}(1997)}]{WiegertHolman1997}
{Wiegert}, P.~A., \& {Holman}, M.~J. 1997, \aj, 113, 1445

\bibitem[{{Wilson}(1978)}]{Wilson1978}
{Wilson}, O.~C. 1978, \apj, 226, 379

\bibitem[{{Wittenmyer} {et~al.}(2014){Wittenmyer}, {Endl}, {Bergmann},
  {Hearnshaw}, {Barnes}, \& {Wright}}]{Wittenmyer2014}
{Wittenmyer}, R.~A., {Endl}, M., {Bergmann}, C., {et~al.} 2014, in IAU
  Symposium, Vol. 293, Formation, Detection, and Characterization of Extrasolar
  Habitable Planets, ed. N.~{Haghighipour}, 58--64

\bibitem[{{Wright}(2005)}]{Wright2005}
{Wright}, J.~T. 2005, \pasp, 117, 657

\end{thebibliography}

\appendix
\section{Radial Velocity Data}
\begin{center}
\begin{longtable}{c c c c c}
\tablenum{2}\\
\caption{Relative, Binned RV Data from the CTIO Telescope}\\
\toprule
 & JD-2440000 & Vel. [m s\textsuperscript{-1}] & Err. [m s\textsuperscript{-1}] & Source \\
\hline
\endfirsthead

\multicolumn{5}{c}%
{{\tablename\ \thetable{} -- continued from previous page}} \\
\toprule
 & JD-2440000 & Vel. [m s\textsuperscript{-1}] & Err. [m s\textsuperscript{-1}] & Source \\
\hline
\endhead

\hline
\multicolumn{5}{r}{{...Continued on next page}} 
\endfoot
\hline
\endlastfoot

\begin{filecontents*}{t-rv.csv}
Target,JD,Vel,Err,Source
A,14689.527,-239.82,7.2,ES
A,14834.8477,-187.13,4.64,ES
A,14835.8433,-178.38,4.85,ES
A,14836.8412,-180.4,5.16,ES
A,14837.8399,-186.24,4.84,ES
A,14838.8282,-185.84,5.04,ES
A,14839.8234,-181.6,4.78,ES
A,14840.8259,-182.87,4.94,ES
A,14841.8235,-185.74,4.89,ES
A,14842.8425,-189.48,5.22,ES
A,14843.833,-180.64,5.15,ES
A,14844.8317,-183.82,5.28,ES
A,14845.8283,-183.83,5.39,ES
A,14846.8287,-188.98,5.31,ES
A,14848.8006,-190.22,5.42,ES
A,14850.8088,-183.51,5.32,ES
A,14851.8037,-182.23,5.62,ES
A,14855.8217,-174.21,5.18,ES
A,14857.8393,-177.05,4.78,ES
A,14862.7957,-175.89,5.06,ES
A,14867.8042,-176.44,5.15,ES
A,14875.7147,-173.45,5.47,ES
A,14881.8382,-168.24,5.55,ES
A,14882.7314,-162.17,5.23,ES
A,14887.7698,-167.12,5.62,ES
A,14888.7879,-163.99,5.57,ES
A,14889.8542,-165.34,5.7,ES
A,14890.7257,-161.29,5.74,ES
A,14894.7156,-160.16,5.43,ES
A,14895.7642,-167.05,5.43,ES
A,14896.6846,-171.9,5.16,ES
A,14901.8234,-164.92,5.15,ES
A,14902.6673,-169.44,4.98,ES
A,14903.7608,-163.29,6.54,ES
A,14904.6589,-162.38,5.12,ES
A,14905.7697,-165.02,4.98,ES
A,14906.6933,-161.75,5.09,ES
A,14907.7316,-164.36,4.94,ES
A,14908.6732,-162.9,5.03,ES
A,14911.6859,-166.83,5.09,ES
A,14912.6442,-166.03,5.05,ES
A,14916.7587,-159.27,5.43,ES
A,14918.7515,-164.44,5.41,ES
A,14919.731,-162.77,5.18,ES
A,14921.8761,-157.52,5.61,ES
A,14922.6839,-160.57,5.33,ES
A,14926.7742,-142.64,4.85,ES
A,14927.694,-147.38,4.93,ES
A,14928.7928,-144.94,4.93,ES
A,14929.7463,-143.53,5.02,ES
A,14930.7301,-144.93,5.03,ES
A,14931.7523,-142.71,5.11,ES
A,14932.7,-141.25,4.96,ES
A,14933.7633,-142.44,5.11,ES
A,14934.6694,-138.23,5.24,ES
A,14935.6033,-138.18,4.95,ES
A,14937.6044,-146.37,5.48,ES
A,14939.7702,-145.74,5.33,ES
A,14940.807,-142.47,5.61,ES
A,14941.7266,-148.14,5.56,ES
A,14942.7769,-140.44,5.9,ES
A,14943.7224,-139.83,5.69,ES
A,14944.8221,-139.98,5.74,ES
A,14945.6187,-135.46,5.27,ES
A,14946.6154,-126.45,5.09,ES
A,14948.6777,-124.19,5.46,ES
A,14949.7201,-124.48,5.44,ES
A,14950.7964,-123.94,5.5,ES
A,14951.7087,-122.32,5.4,ES
A,14952.7778,-120.61,5.86,ES
A,14953.6849,-123.27,5.46,ES
A,14954.5941,-123.33,5.43,ES
A,14961.5962,-115.44,5.75,ES
A,14962.6169,-116.08,5.85,ES
A,14963.6315,-115.07,6.55,ES
A,14964.6743,-109.63,7.41,ES
A,14965.6737,-113.22,5.8,ES
A,14968.6698,-121.15,5.79,ES
A,14969.7843,-131.5,5.54,ES
A,14973.6952,-127.82,5.53,ES
A,14974.6727,-123.47,5.17,ES
A,14975.6759,-124.28,5.18,ES
A,14976.7604,-129.18,5.84,ES
A,14978.6286,-130.09,5.92,ES
A,14982.6575,-129.84,6.04,ES
A,14985.6272,-124.4,6.21,ES
A,14986.6424,-118.12,6.09,ES
A,14987.6358,-117.83,5.94,ES
A,14988.7298,-110.68,5.68,ES
A,14990.6764,-109.05,5.46,ES
A,14991.6645,-107.73,5.51,ES
A,14994.6505,-100.97,5.4,ES
A,14995.5884,-105.16,5.22,ES
A,14998.5407,-98.71,5.84,ES
A,14999.4778,-102.1,7.15,ES
A,15001.6861,-102.45,6.1,ES
A,15004.6872,-98.13,6.14,ES
A,15005.5569,-99.48,5.75,ES
A,15006.5488,-101.72,5.58,ES
A,15007.5487,-100.47,5.81,ES
A,15008.5184,-101.63,5.9,ES
A,15010.5858,-105.65,6.2,ES
A,15020.5845,-102.81,5.47,ES
A,15021.4761,-100.81,5.44,ES
A,15022.6461,-103.75,5.62,ES
A,15023.5657,-106.17,5.35,ES
A,15024.5838,-103.14,5.68,ES
A,15025.5751,-102.23,6.24,ES
A,15028.5416,-103.77,5.74,ES
A,15029.547,-101.93,6.08,ES
A,15030.4899,-99.11,5.88,ES
A,15037.5353,-101.31,5.54,ES
A,15038.5269,-99.41,5.21,ES
A,15039.5346,-99.47,5.19,ES
A,15041.6051,-101.23,4.75,ES
A,15042.5961,-99.27,4.64,ES
A,15045.5327,-95.5,4.71,ES
A,15047.5419,-95.13,4.91,ES
A,15048.5135,-97.97,4.49,ES
A,15049.555,-99.19,5.13,ES
A,15052.5841,-95.34,4.57,ES
A,15071.5221,-93.05,4.71,ES
A,15072.5134,-87.67,4.68,ES
A,15073.5292,-89.76,4.9,ES
A,15275.6896,-5.28,4.98,ES
A,15276.6726,-0.16,5.25,ES
A,15277.7617,-3.98,5.2,ES
A,15280.7366,-5.6,5.03,ES
A,15281.8102,0.99,5.04,ES
A,15282.805,-1.93,5.17,ES
A,15283.769,0.17,5.27,ES
A,15284.7849,-0.15,4.97,ES
A,15286.714,0.15,4.86,ES
A,15287.8027,5.73,4.61,ES
A,15288.7993,-0.99,4.84,ES
A,15289.7368,5.79,4.84,ES
A,15290.7649,8.46,4.55,ES
A,15291.7738,7.63,4.68,ES
A,15292.7165,9.02,5.31,ES
A,15293.7534,14.24,4.73,ES
A,15305.748,16.93,4.93,ES
A,15306.7478,18.49,4.98,ES
A,15307.703,18.09,5.31,ES
A,15308.6617,17.06,5.11,ES
A,15309.8075,22.98,5.23,ES
A,15310.6254,25.2,5.44,ES
A,15312.6215,19.53,5.26,ES
A,15314.7729,35.33,5.93,ES
A,15315.8069,26.05,5.77,ES
A,15316.6916,27.26,5.52,ES
A,15321.7157,25.06,5.57,ES
A,15323.6798,37.18,5.42,ES
A,15325.7765,32.88,5.82,ES
A,15326.6523,33.85,5.24,ES
A,15327.6778,32.94,5.17,ES
A,15334.6222,15.1,5.37,ES
A,15630.7436,159.6,1.38,CHIRON
A,15633.7618,157.82,1.39,CHIRON
A,15634.7746,157.41,1.32,CHIRON
A,15635.7626,154.52,1.3,CHIRON
A,15643.8091,160.2,1.11,CHIRON
A,15646.7974,163.77,1.06,CHIRON
A,15647.7256,160.7,1.05,CHIRON
A,15648.838,159.39,1.07,CHIRON
A,15649.7835,163.53,1.02,CHIRON
A,15650.8004,161.69,1.05,CHIRON
A,15651.7572,161.5,1.18,CHIRON
A,15652.7865,162.04,0.99,CHIRON
A,15653.7891,162.61,1.02,CHIRON
A,15654.7707,164.75,1.02,CHIRON
A,15655.6687,162.16,1.05,CHIRON
A,15656.7466,164.97,1.1,CHIRON
A,15657.7219,164.11,1.11,CHIRON
A,15658.7702,163.59,1.22,CHIRON
A,15659.7116,164.87,1.14,CHIRON
A,15660.6804,168.14,1.16,CHIRON
A,15662.6993,167.53,1.16,CHIRON
A,15663.6964,168.53,1.32,CHIRON
A,15664.6502,165.17,1.31,CHIRON
A,15665.7363,167.0,1.3,CHIRON
A,15670.7739,168.02,1.42,CHIRON
A,16110.5897,352.26,1.39,CHIRON
A,16111.6132,352.03,1.23,CHIRON
A,16113.6057,354.0,1.06,CHIRON
A,16114.5762,354.88,1.17,CHIRON
A,16116.5952,352.1,1.06,CHIRON
A,16117.6046,355.25,1.05,CHIRON
A,16118.5726,356.23,1.34,CHIRON
A,16119.6598,352.54,1.06,CHIRON
A,16120.5393,357.26,0.94,CHIRON
A,16121.6064,357.13,0.97,CHIRON
A,16123.6222,356.18,1.03,CHIRON
A,16124.6205,356.94,1.06,CHIRON
A,16125.5729,347.03,0.88,CHIRON
A,16126.5665,359.25,0.92,CHIRON
A,16127.6297,359.58,1.23,CHIRON
A,16128.6668,360.88,1.11,CHIRON
A,16129.6042,361.17,0.98,CHIRON
A,16130.6275,354.5,0.96,CHIRON
A,16131.5667,363.34,1.04,CHIRON
A,16132.5804,366.17,0.86,CHIRON
A,16133.5668,366.31,1.08,CHIRON
A,16134.5694,365.78,0.9,CHIRON
A,16136.5895,366.19,1.0,CHIRON
A,16137.6584,361.79,1.09,CHIRON
A,16138.6077,369.34,1.02,CHIRON
A,16139.5821,362.26,0.93,CHIRON
A,16140.5724,367.49,0.91,CHIRON
A,16145.6004,367.57,0.94,CHIRON
A,16148.5821,358.02,0.93,CHIRON
A,16149.5451,372.19,0.97,CHIRON
A,16150.463,370.93,1.19,CHIRON
A,16152.4633,367.51,1.01,CHIRON
A,16172.4977,380.39,0.97,CHIRON
A,16176.5071,365.12,0.95,CHIRON
A,16181.4763,387.79,1.04,CHIRON
A,16182.4897,387.2,1.12,CHIRON
A,16183.4983,387.49,1.12,CHIRON
A,16184.4944,383.82,1.14,CHIRON
A,16186.4935,387.8,1.14,CHIRON
A,16190.4802,388.38,0.96,CHIRON
A,16191.4724,383.85,1.12,CHIRON
A,16192.4572,393.27,1.22,CHIRON
A,16193.4678,389.3,1.1,CHIRON
A,16194.4879,388.88,1.23,CHIRON
A,16195.4686,392.29,1.06,CHIRON
A,16196.481,390.46,0.93,CHIRON
A,16197.4705,387.74,0.9,CHIRON
A,16199.5065,391.51,1.03,CHIRON
A,16200.4903,387.37,1.2,CHIRON
B,14834.835,119.94,4.05,ES
B,14835.8154,126.49,4.74,ES
B,14844.8337,122.98,5.06,ES
B,14845.7948,119.91,4.32,ES
B,14862.7934,124.01,4.36,ES
B,14880.7889,100.79,4.29,ES
B,14881.7848,101.0,4.11,ES
B,14882.7524,108.03,4.51,ES
B,14883.7437,105.58,4.14,ES
B,14884.7105,102.26,4.3,ES
B,14887.7151,117.18,5.24,ES
B,14894.7039,109.19,4.84,ES
B,14895.7642,98.88,4.31,ES
B,14896.6933,100.91,4.09,ES
B,14897.7612,107.82,4.88,ES
B,14898.6603,102.91,4.32,ES
B,14899.7246,103.61,4.09,ES
B,14900.6952,103.72,4.43,ES
B,14901.7594,97.89,4.49,ES
B,14902.6592,94.33,3.94,ES
B,14903.7576,103.24,5.94,ES
B,14904.6592,95.87,3.97,ES
B,14905.7776,92.31,4.31,ES
B,14906.6932,90.59,4.2,ES
B,14907.7309,91.06,4.21,ES
B,14908.7252,90.66,4.51,ES
B,14909.7232,85.04,4.57,ES
B,14911.6886,84.08,4.9,ES
B,14912.7551,79.57,4.44,ES
B,14913.6642,75.52,4.16,ES
B,14914.6519,81.14,4.16,ES
B,14916.7501,82.35,4.67,ES
B,14918.7581,74.47,3.97,ES
B,14921.8821,76.29,4.5,ES
B,14922.7541,72.19,4.57,ES
B,14923.7507,72.54,4.12,ES
B,14924.5908,71.22,3.81,ES
B,14929.7464,74.33,4.57,ES
B,14930.6643,65.46,4.4,ES
B,14934.6838,69.64,4.53,ES
B,14935.6539,69.19,4.31,ES
B,14939.7635,61.16,4.1,ES
B,14940.7127,71.12,4.64,ES
B,14944.8196,68.99,4.52,ES
B,14945.5655,69.94,4.21,ES
B,14948.6814,78.91,4.52,ES
B,14949.635,76.89,4.31,ES
B,14953.6845,58.26,4.1,ES
B,14954.6303,58.35,4.4,ES
B,14955.6246,63.46,5.99,ES
B,14961.6295,69.52,4.97,ES
B,14962.6319,69.0,5.29,ES
B,14963.64,69.32,6.14,ES
B,14964.6822,67.66,6.89,ES
B,14965.6801,63.58,4.9,ES
B,14972.6819,60.96,4.78,ES
B,14973.6822,62.35,4.73,ES
B,14974.6726,59.86,4.96,ES
B,14975.6761,60.38,4.65,ES
B,14976.7605,52.06,5.35,ES
B,14978.6361,47.82,6.63,ES
B,14980.6632,50.99,8.91,ES
B,14982.6316,47.7,5.04,ES
B,14985.6196,53.18,4.88,ES
B,14986.5313,49.78,4.82,ES
B,14990.6837,43.67,4.7,ES
B,14991.6639,37.09,4.94,ES
B,14994.6473,35.89,4.45,ES
B,14995.6165,36.06,4.38,ES
B,14998.5302,38.52,4.68,ES
B,15020.576,30.26,4.41,ES
B,15021.4787,32.42,4.88,ES
B,15022.5835,35.97,4.55,ES
B,15023.5364,33.1,4.55,ES
B,15024.5759,34.67,4.98,ES
B,15025.5884,39.24,5.49,ES
B,15028.5414,36.77,5.14,ES
B,15029.547,32.26,4.78,ES
B,15030.5056,27.94,4.54,ES
B,15032.4931,26.66,4.29,ES
B,15036.5311,19.32,4.47,ES
B,15037.5274,18.48,4.64,ES
B,15038.5273,20.24,4.35,ES
B,15039.5271,19.54,4.35,ES
B,15041.5576,17.13,3.48,ES
B,15042.5438,15.78,3.59,ES
B,15045.5424,21.53,3.59,ES
B,15047.5527,21.51,3.5,ES
B,15049.5654,21.83,3.8,ES
B,15052.5741,17.99,3.84,ES
B,15053.5395,13.52,3.82,ES
B,15054.5274,13.45,3.72,ES
B,15055.4897,8.85,5.33,ES
B,15057.54,13.61,3.89,ES
B,15058.5289,15.54,4.14,ES
B,15067.5265,10.13,3.65,ES
B,15068.5128,8.61,3.81,ES
B,15069.5023,8.72,4.02,ES
B,15070.5333,2.58,3.87,ES
B,15275.6936,-102.57,3.61,ES
B,15276.6908,-98.43,3.6,ES
B,15280.73,-108.2,3.63,ES
B,15281.8107,-101.61,3.61,ES
B,15282.8236,-109.26,3.67,ES
B,15283.77,-107.68,3.61,ES
B,15284.8214,-108.63,3.48,ES
B,15286.7174,-109.06,3.52,ES
B,15287.8028,-103.87,3.65,ES
B,15288.8174,-116.0,3.59,ES
B,15289.7415,-118.11,3.81,ES
B,15290.7836,-117.32,3.55,ES
B,15291.793,-117.06,3.45,ES
B,15292.7347,-119.47,3.86,ES
B,15293.79,-115.51,3.66,ES
B,15305.7671,-114.92,3.58,ES
B,15306.7659,-110.84,3.52,ES
B,15307.7214,-110.74,3.72,ES
B,15308.6617,-110.61,3.43,ES
B,15309.8075,-102.39,3.62,ES
B,15310.672,-102.47,4.21,ES
B,15312.656,-101.03,3.72,ES
B,15314.7729,-101.2,3.93,ES
B,15315.8268,-102.63,4.31,ES
B,15316.7134,-110.98,4.06,ES
B,15321.7316,-126.63,4.23,ES
B,15323.6764,-133.55,3.89,ES
B,15325.715,-124.68,4.02,ES
B,15326.6822,-123.64,4.21,ES
B,15327.7727,-126.18,4.08,ES
B,15334.5725,-142.84,4.26,ES
B,15654.8333,-299.67,1.11,CHIRON
B,15655.8039,-299.18,1.08,CHIRON
B,15656.8657,-293.14,1.08,CHIRON
B,15657.7894,-301.94,1.04,CHIRON
B,15658.8223,-299.75,1.14,CHIRON
B,15659.8,-301.51,1.03,CHIRON
B,15660.786,-301.63,1.07,CHIRON
B,15662.8428,-305.33,1.01,CHIRON
B,15665.7569,-289.88,1.2,CHIRON
B,15667.6995,-312.13,1.01,CHIRON
B,15669.7492,-299.48,1.14,CHIRON
B,15670.7316,-321.56,1.15,CHIRON
B,15671.8097,-301.57,1.15,CHIRON
B,15674.8066,-311.48,1.19,CHIRON
B,15685.8156,-342.6,1.52,CHIRON
B,16110.6054,-549.45,1.34,CHIRON
B,16111.6467,-534.59,1.12,CHIRON
B,16113.621,-550.75,1.01,CHIRON
B,16114.5529,-546.26,1.03,CHIRON
B,16116.6095,-544.78,1.08,CHIRON
B,16117.5873,-534.15,1.09,CHIRON
B,16118.5623,-552.82,1.22,CHIRON
B,16119.6689,-538.98,1.19,CHIRON
B,16120.5583,-535.16,0.97,CHIRON
B,16121.628,-532.54,0.97,CHIRON
B,16123.6302,-546.11,1.16,CHIRON
B,16124.6638,-531.36,0.98,CHIRON
B,16126.5825,-534.37,1.0,CHIRON
B,16127.6112,-551.61,1.07,CHIRON
B,16128.6924,-546.33,1.27,CHIRON
B,16129.6014,-554.56,1.25,CHIRON
B,16131.5852,-547.99,1.15,CHIRON
B,16133.5813,-559.88,1.14,CHIRON
B,16134.5919,-556.91,1.1,CHIRON
B,16136.6106,-558.03,1.02,CHIRON
B,16137.6507,-556.03,1.1,CHIRON
B,16138.6185,-563.05,1.05,CHIRON
B,16139.602,-556.03,1.02,CHIRON
B,16140.587,-560.25,1.02,CHIRON
B,16145.5771,-564.0,1.02,CHIRON
B,16172.5137,-575.94,1.01,CHIRON
B,16181.4911,-583.37,1.04,CHIRON
B,16182.5059,-584.77,1.05,CHIRON
B,16183.511,-584.12,1.14,CHIRON
B,16186.5042,-588.47,1.22,CHIRON
B,16190.5024,-584.27,1.14,CHIRON
B,16191.4872,-590.68,1.2,CHIRON
B,16193.4826,-588.77,1.13,CHIRON
B,16194.5008,-591.73,1.2,CHIRON
B,16195.479,-593.01,1.14,CHIRON
B,16196.4945,-589.53,1.19,CHIRON
B,16197.484,-587.77,1.04,CHIRON
B,16199.4924,-581.69,1.53,CHIRON
B,16200.5106,-596.93,1.42,CHIRON
B,16204.4966,-586.71,1.33,CHIRON
B,16306.8184,-656.56,1.12,CHIRON
B,16307.8838,-660.37,1.28,CHIRON
B,16308.7993,-662.78,1.17,CHIRON
B,16311.8617,-658.04,1.19,CHIRON
B,16312.8287,-658.31,1.29,CHIRON
B,16314.8736,-645.95,1.09,CHIRON
B,16316.8335,-638.73,1.09,CHIRON
B,16318.7593,-666.39,1.05,CHIRON
B,16320.8406,-662.46,1.05,CHIRON
B,16321.8254,-665.56,1.07,CHIRON
B,16322.8044,-659.49,1.05,CHIRON
B,16323.8111,-671.61,1.09,CHIRON
B,16324.8839,-658.87,1.06,CHIRON
B,16326.8693,-666.14,1.01,CHIRON
B,16327.855,-667.73,1.08,CHIRON
B,16330.8422,-656.82,1.18,CHIRON
B,16332.8404,-648.0,1.02,CHIRON
B,16334.8317,-662.86,1.04,CHIRON
B,16335.8603,-656.0,1.08,CHIRON
B,16336.8252,-659.85,1.1,CHIRON
B,16338.8316,-674.75,1.04,CHIRON
B,16340.8571,-670.17,1.02,CHIRON
B,16342.8366,-665.08,1.03,CHIRON
B,16344.6894,-647.81,1.02,CHIRON
B,16384.8077,-696.29,0.99,CHIRON
B,16385.7757,-693.22,1.16,CHIRON
B,16386.7955,-702.42,1.02,CHIRON
B,16387.7926,-701.36,1.02,CHIRON
B,16388.8209,-699.19,1.06,CHIRON
B,16389.7882,-697.77,1.1,CHIRON
B,16390.7706,-709.39,1.08,CHIRON
B,16391.7455,-709.97,1.08,CHIRON
B,16395.791,-693.25,1.15,CHIRON
B,16398.7314,-678.64,1.22,CHIRON
B,16400.7341,-687.89,0.97,CHIRON
B,16401.7961,-700.19,1.27,CHIRON
B,16403.571,-707.51,1.16,CHIRON
B,16404.7147,-707.2,1.15,CHIRON
B,16405.5749,-709.89,7.39,CHIRON
B,16406.7852,-713.0,1.25,CHIRON
B,16407.7163,-695.09,1.3,CHIRON
B,16415.8016,-705.04,1.04,CHIRON
B,16417.6262,-701.56,0.97,CHIRON
B,16418.7307,-725.92,1.3,CHIRON
B,16419.6953,-721.54,1.16,CHIRON
B,16420.6911,-720.32,1.2,CHIRON
B,16421.5491,-714.46,1.52,CHIRON
B,16423.7005,-718.31,1.13,CHIRON
B,16424.761,-726.43,1.44,CHIRON
B,16435.6685,-729.09,1.05,CHIRON
B,16436.675,-708.93,1.12,CHIRON
B,16437.6636,-717.18,1.11,CHIRON
\end{filecontents*}
\csvreader[late after line=\\, column count=5]
{t-rv.csv}
{Target=\Target, JD=\JD, Vel=\Vel, Err=\Err, Source=\Source}
{\Target & \JD & \Vel & \Err & \Source}
\end{longtable}
\end{center}
\end{document}